\documentclass[]{emulateapj}

\bibpunct{(}{)}{;}{a}{}{,}

\slugcomment{Accepted for publication in ApJ}
\shorttitle{The chemical evolution of Bo\"otes\,I}
\shortauthors{Frebel et al.} 

\begin{document}

\title{The chemical evolution of the Bo\"otes\,I ultra-faint dwarf
  galaxy\altaffilmark{*}}

\author{
Anna Frebel\altaffilmark{1,2},
John E. Norris\altaffilmark{3}, 
Gerard Gilmore\altaffilmark{4},
Rosemary F. G. Wyse\altaffilmark{5}}

\altaffiltext{*}{This paper includes data gathered with the 6.5 meter
  Magellan Telescopes located at Las Campanas Observatory, Chile.}

\altaffiltext{1}{Department of Physics and Kavli Institute for
  Astrophysics and Space Research, Massachusetts Institute of
  Technology, Cambridge, MA 02139, USA}

\altaffiltext{2}{Joint Institute for Nuclear Astrophysics - Center for Evolution of the Elements, East Lansing, MI 48824}

\altaffiltext{3}{Research School of Astronomy and Astrophysics,
  Australian National University, Canberra, ACT, 2611, Australia}

\altaffiltext{4}{Institute of Astronomy, University of Cambridge, 
 Cambridge CB3 0HA, UK}

\altaffiltext{5}{The Johns Hopkins University, Department of Physics \&
  Astronomy,  Baltimore, MD 21218, USA}

\begin{abstract}

We present chemical abundance measurements of two metal-poor
  red giant stars in the ultra-faint dwarf galaxy Bo\"otes\,I, based
  on Magellan/MIKE high-resolution spectra. For Boo\,I-980, with
  $\mbox{[Fe/H]}=-3.1$, we present the first elemental abundance
  measurements while Boo\,I-127, with $\mbox{[Fe/H]}=-2.0$, shows
  abundances in good agreement with previous measurements.  Light and
  iron-peak element abundance ratios in the two Bo\"otes\,I stars, as
  well as those of most other Bo\"otes\,I members, collected from the
  literature, closely resemble those of regular metal-poor halo stars.
  Neutron-capture element abundances Sr and Ba are systematically
  lower than the main halo trend, and also show a significant
  abundance spread. Overall, this is similar to what has been found
  for other ultra-faint dwarf galaxies. We apply corrections to the
  carbon abundances (commensurate with stellar evolutionary status) of
  the entire sample and find 21\% of stars to be carbon-enhanced
  metal-poor (CEMP) stars, compared to 13\% without using the carbon
  correction. We reassess the metallicity distribution functions (MDF)
  for the CEMP stars and non-CEMP stars, and confirm earlier claims
  that CEMP stars might belong to a different, earlier
  population. Applying a set of abundance criteria to test to what
  extent Bo\"otes\,I could be a surviving first galaxy suggests that
  it is one of the earliest assembled systems that perhaps received
  gas from accretion from other clouds in the system, or
  from swallowing a first galaxy or building block type object. This
  resulted in the two stellar populations observable today.  
\end{abstract}


\section{Introduction}

Dwarf satellite galaxies are versatile probes of chemical evolution,
galactic halo assembly, and early galaxy formation processes.
Ultra-faint dwarf galaxies (L$<10^{5}$\,L$_{\odot}$) are particularly
well-suited for testing these processes since they appear to have had
limited star formation and chemical evolution, thus rendering them relatively
simple and unevolved systems. What has made ultra-faints particularly
interesting is their overall low-metallicity coupled with large
metallicity spreads of several dex (e.g., \citep{kirby08}). While
stars with metallicities below $\mbox{[Fe/H]}=-3.0$ have been found in
nearly all of these systems, the ultra-faint dwarfs completely lack
higher-metallicity stars $\mbox{[Fe/H]}>-1.0$ showing that chemical
enrichment did not proceed long enough to reach even close to the
solar level.

Based on the chemical abundances of the seven brightest stars, the
faintest of the ultra-faint dwarfs, Segue\,1, was found to be a
promising candidate for a surviving first galaxy
\citep{frebel14}. Criteria for such a survivor include large spreads
in [Fe/H], halo-like $\alpha$-abundances at higher metallicity and low
neutron-capture element abundances, all of which are found among
Segue\,1 stars. Being the faintest of all ultra-faint dwarf galaxies,
Segue\,1 may be regarded the best candidate for one of the most
primitive dwarf galaxies still observable (see also \citealt{ji15} and
\citealt{webster15}). But how common could surviving first, or very
early, galaxies be today? This can only be answered with detailed
inspections of the stellar chemical abundances in additional dwarf
systems. This is, however, challenging given the faint nature of even
the brightest available stars near the tip of the red giant
branch. High-resolution spectra are thus only available of few stars
per galaxy (e.g., three each in Ursa Major\,II and Coma Berenices:
\citealt{frebel10}; one star in Leo\,IV: \citealt{leo4}; twelve stars
in Hercules although most of them have a very limited red wavelength
coverage: \citealt{koch_her, aden11}; seven stars in Segue\,1:
\citealt{norris10_seg,frebel14}; four stars on Bo\"otes\,II
\citealt{koch14}; \citealt{ji16a}).

Fortunately, Bo\"otes\,I has been studied rather extensively since its
discovery in 2006 by \citet{belokurov06}. Beginning with
medium-resolution spectroscopic surveys \citep{munoz06,norris_boo,
  norris10_booseg}, membership was established and then [Fe/H]
distributions. \citet{hughes08,hughes14} also find a large abundance
spread of at least from $-3.7 < \mbox{[Fe/H]} < -1.9$ based on
photometric metallicities. [C/Fe] spreads and $\alpha$-abundance were
also spectroscopically measured for some subsamples
\citep{lai11}. From these samples, a total of 11 stars have been
observed at high-resolution to produce chemical abundance patterns of
these stars \citep{feltzing09, norris10, gilmore13, ishigaki14}, of
which most have been observed by more than one group. This has led to
well-established abundances for these stars, which is of importance
for the overall interpretation of the chemical inventory of
Bo\"otes\,I. Additionally, a kinematic study \citet{koposov11} has
been carried out, age estimates Bo\"otes\,I were obtained
\citep{okamoto12,hughes14,brown14b} and its chemical evolution modeled
\citep{romano15}.

In this paper we report observations for one star in Bo\"otes\,I with
no previous high-resolution spectroscopy, and a second that has been
observed before. Our goal is to investigate the chemical enrichment
history of Bo\"otes\,I to find out to what extent this galaxy
  shows signs of a surviving first galaxy. To test such a hypothesis,
  we use the observational criteria suggested by \citet{frebel12}
  about the metallicity spread and different abundance levels in a
  given system. As part of this we also aim at quantifying whether
there were multiple stellar generations present and to what extent
chemical \textit{evolution} occurred in this system. This
  follows earlier hints that some element contributions by supernovae
  type Ia have occurred in the system \citep{gilmore13}. Since the
luminosity of Bo\"otes\,I is about 100 times higher than that of
Segue\,1, and nearly 10$^5$\,L$_{\odot}$ \citep{munoz06}, we also aim
at addressing how the overall extent of chemical evolution relates to
the luminosity of the system. Additional considerations besides the
metallicity-luminosity relationship (e.g., \citealt{kirby08}) that
appears to exist for all dwarf galaxies may shed light on the
formation and evolution of these objects.

In \S\,\ref{sec:obs} we describe our new observations and in
\S\,\ref{sec:analysis} our analysis techniques and corresponding
chemical abundance results. We apply a set of chemical abundances
criteria to our and literature abundances for assessing to what extent
Bo\"otes\,I is a surviving first galaxy in \S\,\ref{signature}. We
discuss our findings in \S\,\ref{sec:conc}.

\section{Target Selection and Observations}\label{sec:obs}

Targets were selected from \citet{norris_boo} who had taken medium
resolution $R\sim5,000$ AAOmega spectra of Bo\"otes\,I stars with the
Anglo-Australian Telescope (AAT). Their wide survey covered $\sim6$
half-light radii from the center of the galaxy.  Radial
velocity-confirmed members were further analyzed
\citep{norris10_booseg}. Star Boo-1137, with $\mbox{[Fe/H]}=-3.7$, was
observed with high spectral resolution for a detailed abundance
analysis \citep{norris10}, as were seven additional stars
\citep{gilmore13}. Six of those stars were also observed by
\citet{ishigaki14}.  The only brighter star that had not been observed
was Boo-980. Being located at 3.9 half-light radii from the center it
was only covered by the \citet{norris10_booseg} study. On the
contrary, another star, Boo-127, was observed by all these studies,
first by \citet{feltzing09} and then by all others. Discordant
abundance ratios were found which made this star interesting for
re-observation. We observed these latter two Bo\"otes\,I stars with
the MIKE spectrograph \citep{mike} on the Magellan-Clay telescope in
March 2010 and March 2011. Details of the MIKE observations and
photometry taken from Norris et al. (2008) are given in
Table~\ref{Tab:obs}.  MIKE spectra have nearly full optical wavelength
coverage over the range of $\sim3500$-9000\,{\AA}.  Observing
conditions during these runs were mostly clear, with an average seeing
of 0.8\arcsec\ to 1.0\arcsec.  The $1.0\arcsec\ \times 5\arcsec$ slit
yields a spectral resolution of $\sim22,000$ in the red and
$\sim28,000$ in the blue wavelength regime. We used $2\times2$ on-chip
binning.  Exposure times were $\sim4$ and 7\,hr, and the observations
were typically broken up into 45 to 55\,min exposures to avoid
significant degradation of the spectra by cosmic rays.

\begin{deluxetable*}{lrrcccrccccc} 
\tablewidth{0pt} 
\tabletypesize{\scriptsize}
\tablecaption{\label{Tab:obs} Observing Details }
\tablehead{
\colhead{Star} & \colhead{$\alpha$}&\colhead{$\delta$}&\colhead{UT dates}&\colhead{slit}&\colhead{$t_{\rm {exp}}$} 
&\colhead{$g_{0}$} &\colhead{$(g-r)_{0}$} &\colhead{$S/N$} &\colhead{$S/N$}   \\
\colhead{}&\colhead{(J2000)}&\colhead{(J2000)}&\colhead{ }&\colhead{}&\colhead{hr}&\colhead{mag}&\colhead{mag}&\colhead{5300\,{\AA}}&\colhead{6000\,{\AA}} }
\startdata
SDSS\,JBoo980& 13 59 12.68 &+13 42 55.8  & 2010-03-07/08/18/24 &$1.0\arcsec$ & 7.0& 18.51& 0.61 & 25& 37\\ 
SDSS\,JBoo127& 14 00 14.57 &+14 35 52.7  & 2011-03-10/12       &$1.0\arcsec$ & 3.7& 18.15& 0.76 & 30& 43\\ 
\enddata
\tablecomments{The $S/N$ is measured per $\sim33$\,m{\AA} pixel.}
\end{deluxetable*}

\begin{figure*}[!t]
 \begin{center}
  \includegraphics[clip=true,width=8cm, bbllx=44, bblly=124,
    bburx=558, bbury=598]{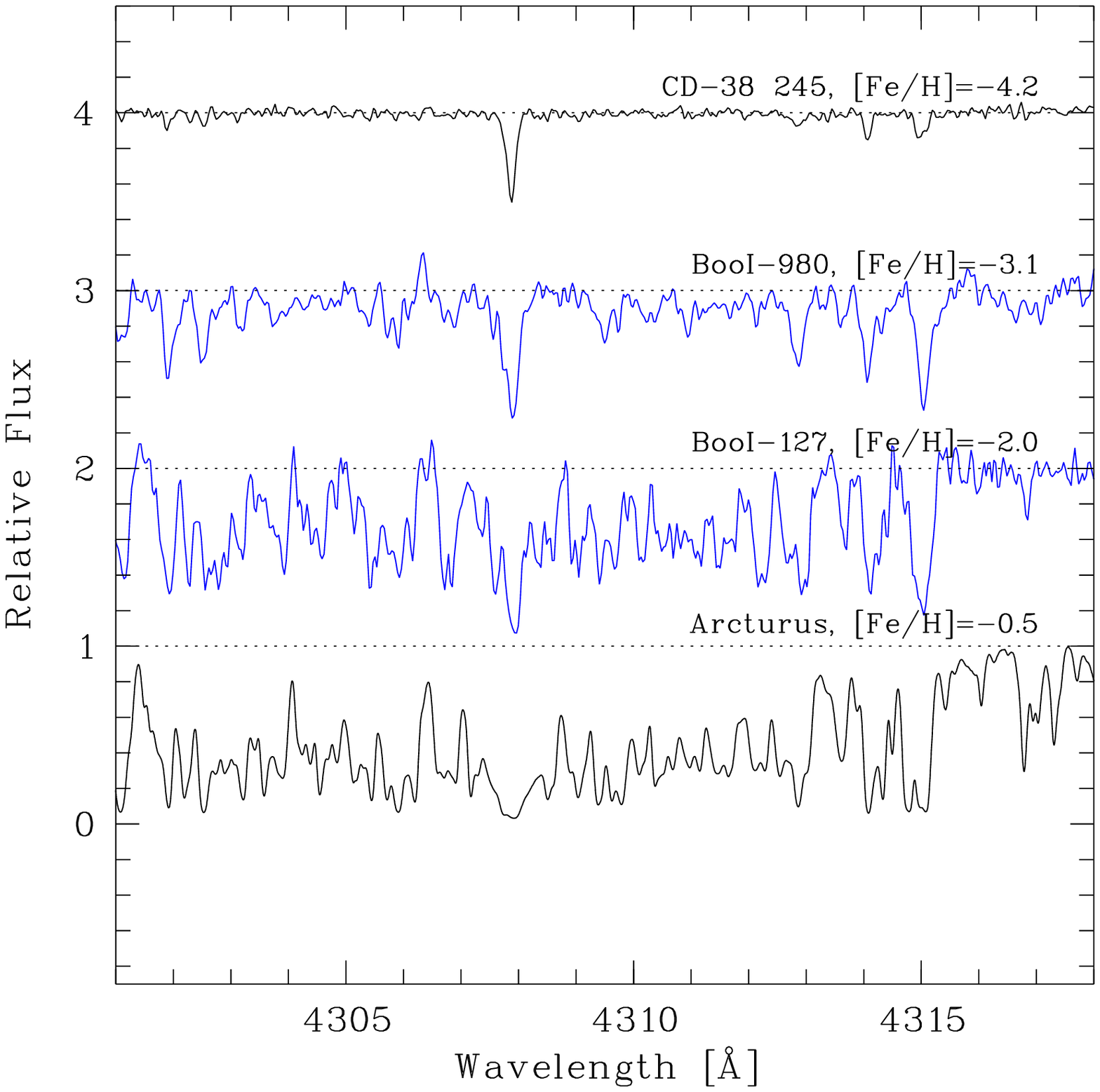}
  \includegraphics[clip=true,width=8cm, bbllx=44, bblly=124,
    bburx=558, bbury=598]{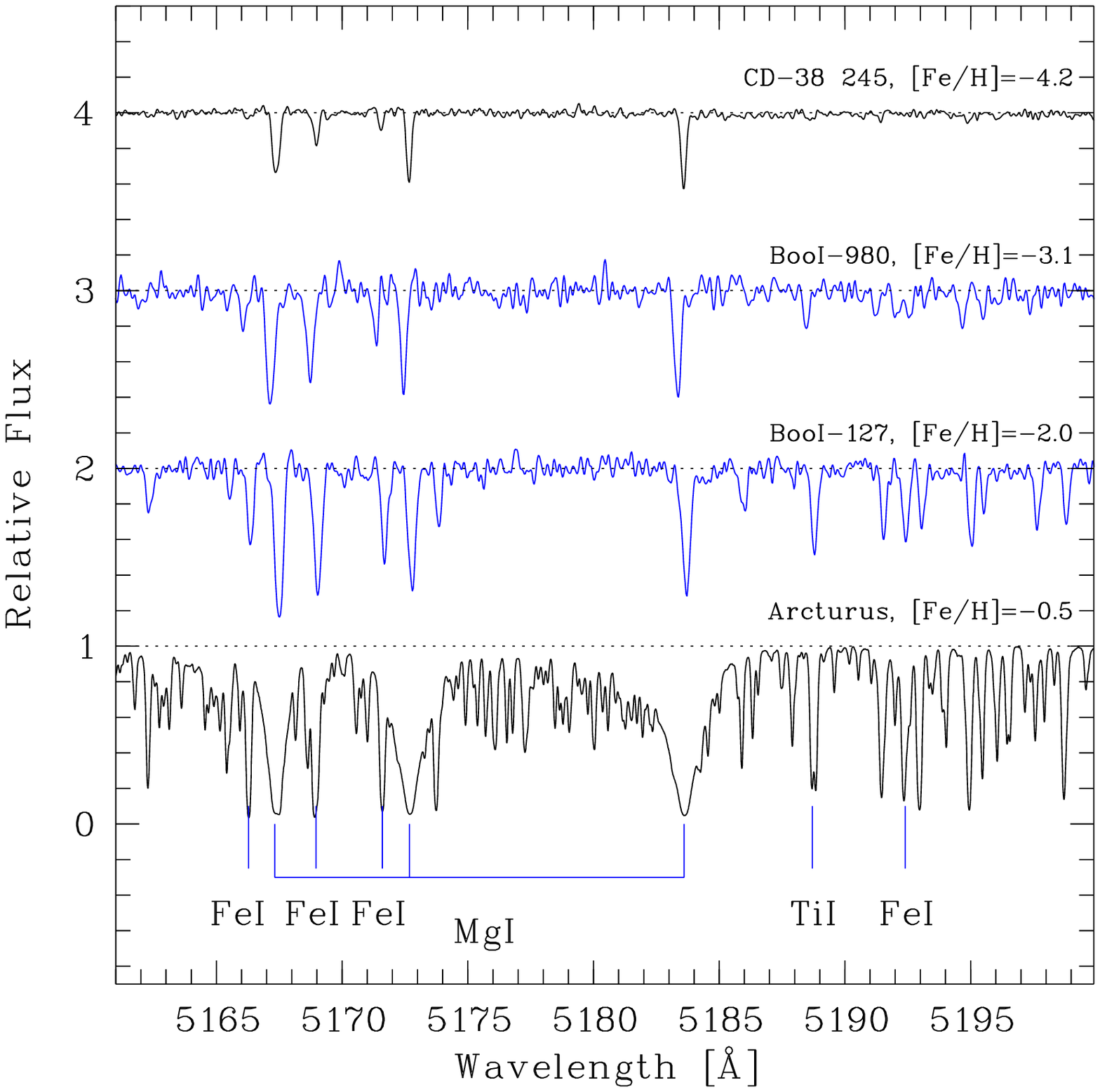}
  \figcaption{\label{specs}Magellan/MIKE spectra of our Bo\"otes\,I
    stars, shown near the G-band head at 4314\,{\AA} (left) and the
    Mg\,b lines around 5180\,{\AA} (right).  Some absorption lines are
    indicated.  The stars are bracketed in terms of their metallicity
    by Arcturus (bottom) and CD~$-$38~245 (top). }
 \end{center}
\end{figure*}

Reductions of the individual MIKE spectra were carried out with the
MIKE Carnegie Python pipeline initially described by
\citet{kelson03}\footnote{Available at
  http://obs.carnegiescience.edu/Code/python}.  The orders of the
combined spectrum were normalized and merged to produce final
one-dimensional blue and red spectra for further analysis. The spectra
have modest $S/N$ which ranges from 25 to 30 at $\sim5300$\,{\AA} and
37 to 43 at $\sim6000$\,{\AA}. Figure~\ref{specs} shows a
representative portions of the spectra around the CH G-band at
4313\,{\AA} and the Mg\,b lines at 5170\,{\AA}. For comparison we also
add CD~$-$38~245 with $\mbox{[Fe/H]}\sim-4.2$ as well as Arcturus
($\mbox{[Fe/H]}=-0.5$). They bracket the metallicities of the
Bo\"otes\,I stars.

\section{Chemical abundance analysis results}\label{sec:analysis}

We measured the equivalent widths of metal absorption lines throughout
the spectra using software from \citet{casey14}. Blending features
were treated with spectrum synthesis to obtain line
abundances. Table~\ref{Tab:Eqw} lists the lines used and their
measured equivalent widths and abundances for all elements, together
with 3$\sigma$ upper limits for selected elements. Full details of our
analysis procedure are given in \citet{frebel14} who analyzed the
Segue\,1 stars in exactly the same manner.

\begin{deluxetable*}{lllrrrrr}
\tablecolumns{8}
\tablewidth{0pt}
\tabletypesize{\footnotesize}
\tabletypesize{\tiny}
\tablecaption{\label{Tab:Eqw} Equivalent width measurements of the Bo\"otes\,I stars  (Table will be available electronically)}
\tablehead{
\colhead{El.} & \colhead{$\lambda$} &\colhead{$\chi$} &\colhead{$\log gf$}&\colhead{EW [m{\AA}]}&\colhead{$\lg\epsilon$ [dex]}&
\colhead{EW [m{\AA}]}&\colhead{$\lg\epsilon$ [dex]}\\
\colhead{}&\colhead{[{\AA}]}&\colhead{[eV]}&\colhead{[dex]}&
\multicolumn{2}{c}{Boo-980}& 
\multicolumn{2}{c}{Boo-127}   \\
}
\startdata
 CH    &   4313     &  \nodata  &  \nodata  &    syn   &      4.97  &    syn   &      6.24 \\ 
 CH    &   4323     &  \nodata  &  \nodata  &    syn   &      4.97  &    syn   &      6.34 \\ 
 Na\,I &   5889.950 &      0.00 &     0.108 &    153.6 &      3.56  &    218.6 &      4.19 \\
 Na\,I &   5895.924 &      0.00 &  $-$0.194 &    129.9 &      3.46  &    199.4 &      4.27 \\
 Mg\,I &   3829.355 &      2.71 &  $-$0.208 &    163.7 &      4.96  &   \nodata&   \nodata \\
 Mg\,I &   3838.292 &      2.72 &     0.490 &    199.2 &      4.61  &   \nodata&   \nodata \\
 Mg\,I &   3986.753 &      4.35 &  $-$1.030 &   \nodata&   \nodata  &     88.7 &      6.16 \\
 Mg\,I &   4057.505 &      4.35 &  $-$0.890 &   \nodata&   \nodata  &     96.3 &      6.15 \\
 Mg\,I &   4167.271 &      4.35 &  $-$0.710 &   \nodata&   \nodata  &     97.7 &      5.99 \\
 Mg\,I &   4571.096 &      0.00 &  $-$5.688 &     46.2 &      4.92  &    122.6 &      6.06 \\
 Mg\,I &   4702.990 &      4.33 &  $-$0.380 &     44.3 &      4.73  &    117.9 &      5.87 \\
 Mg\,I &   5172.684 &      2.71 &  $-$0.450 &    166.3 &      4.90  &   \nodata&   \nodata \\
 Mg\,I &   5183.604 &      2.72 &  $-$0.239 &    169.5 &      4.73  &   \nodata&   \nodata \\
 Mg\,I &   5528.405 &      4.34 &  $-$0.498 &     49.8 &      4.94  &    114.5 &      5.91 \\
 Al\,I &   3944.006 &      0.00 &  $-$0.644 &     88.4 &      2.42  &   \nodata&   \nodata \\
 Al\,I &   3961.520 &      0.01 &  $-$0.340 &    119.2 &      2.81  &    171.9 &      3.57 \\
 Si\,I &   3905.523 &      1.91 &  $-$1.092 &    167.6 &      4.87  &   \nodata&   \nodata \\
 Si\,I &   4102.936 &      1.91 &  $-$3.140 &     63.3 &      4.90  &    138.7 &      6.28 \\
 Ca\,I &   4226.730 &      0.00 &     0.244 &    183.4 &      3.47  &   \nodata&   \nodata \\
 Ca\,I &   4283.010 &      1.89 &  $-$0.224 &   \nodata&   \nodata  &    107.4 &      4.75 \\
 Ca\,I &   4318.650 &      1.89 &  $-$0.210 &   \nodata&   \nodata  &    100.4 &      4.59 \\
 Ca\,I &   4425.440 &      1.88 &  $-$0.358 &     38.6 &      3.61  &     98.0 &      4.65 \\
 Ca\,I &   4434.960 &      1.89 &  $-$0.010 &     53.9 &      3.55  &   \nodata&   \nodata \\
 Ca\,I &   4435.690 &      1.89 &  $-$0.519 &   \nodata&   \nodata  &     96.9 &      4.79 \\
 Ca\,I &   4454.780 &      1.90 &     0.260 &     60.9 &      3.42  &    124.4 &      4.56 \\
 Ca\,I &   4455.890 &      1.90 &  $-$0.530 &     31.6 &      3.67  &   \nodata&   \nodata \\
 Ca\,I &   5581.971 &      2.52 &  $-$0.555 &   \nodata&   \nodata  &     38.0 &      4.50 \\
 Ca\,I &   5588.760 &      2.52 &     0.210 &   \nodata&   \nodata  &     84.1 &      4.49 \\
 Ca\,I &   5590.120 &      2.52 &  $-$0.571 &   \nodata&   \nodata  &     38.3 &      4.51 \\
 Ca\,I &   5594.468 &      2.52 &     0.097 &   \nodata&   \nodata  &     86.8 &      4.64 \\
 Ca\,I &   5598.487 &      2.52 &  $-$0.087 &   \nodata&   \nodata  &     73.6 &      4.61 \\
 Ca\,I &   5601.285 &      2.53 &  $-$0.523 &   \nodata&   \nodata  &     34.6 &      4.41 \\
 Ca\,I &   5857.450 &      2.93 &     0.230 &   \nodata&   \nodata  &     74.9 &      4.79 \\
 Ca\,I &   6102.720 &      1.88 &  $-$0.790 &   \nodata&   \nodata  &     86.2 &      4.70 \\
 Ca\,I &   6122.220 &      1.89 &  $-$0.315 &     36.9 &      3.47  &    107.5 &      4.59 \\
 Ca\,I &   6162.170 &      1.90 &  $-$0.089 &     63.0 &      3.70  &    133.8 &      4.79 \\
 Ca\,I &   6169.055 &      2.52 &  $-$0.797 &   \nodata&   \nodata  &     34.7 &      4.66 \\
 Ca\,I &   6169.559 &      2.53 &  $-$0.478 &   \nodata&   \nodata  &     48.7 &      4.59 \\
 Ca\,I &   6439.070 &      2.52 &     0.470 &     53.6 &      3.71  &     93.3 &      4.33 \\
 Ca\,I &   6449.810 &      2.52 &  $-$0.502 &   \nodata&   \nodata  &     41.6 &      4.47 \\
 Ca\,I &   6499.649 &      2.52 &  $-$0.818 &   \nodata&   \nodata  &     36.1 &      4.69 \\
 Ca\,I &   6717.685 &      2.71 &  $-$0.524 &   \nodata&   \nodata  &     29.3 &      4.48 \\
Sc\,II &   4246.820 &      HFS  & \nodata   &    116.7 &      0.19  &      syn &      1.01 \\
Sc\,II &   4314.083 &      HFS  & \nodata   &     74.6 &   $-$0.04  &      syn &      1.14 \\
Sc\,II &   4324.998 &      HFS  & \nodata   &     59.0 &   $-$0.03  &      syn &      1.15 \\
Sc\,II &   4400.389 &      HFS  & \nodata   &     59.8 &      0.11  &      syn &      1.15 \\ 
Sc\,II &   4415.544 &      HFS  & \nodata   &     56.7 &      0.16  &      syn &      1.05 \\
Sc\,II &   5031.010 &      HFS  & \nodata   &   \nodata&   \nodata  &      syn &      1.00 \\
Sc\,II &   5239.811 &      HFS  & \nodata   &   \nodata&   \nodata  &      syn &      1.00 \\
Sc\,II &   5526.785 &      HFS  & \nodata   &     18.8 &      0.09  &      syn &      1.00 \\
Sc\,II &   5657.907 &      HFS  & \nodata   &   \nodata&   \nodata  &      syn &      1.13 \\
 Ti\,I &   3989.760 &      0.02 &  $-$0.062 &     74.5 &      2.34  &   \nodata&   \nodata \\
 Ti\,I &   3998.640 &      0.05 &     0.010 &     72.8 &      2.27  &   \nodata&   \nodata \\
 Ti\,I &   4512.730 &      0.84 &  $-$0.424 &   \nodata&   \nodata  &     45.4 &      3.06 \\
 Ti\,I &   4518.020 &      0.83 &  $-$0.269 &     26.7 &      2.55  &     49.6 &      2.97 \\
 Ti\,I &   4533.249 &      0.85 &     0.532 &     40.5 &      2.05  &     94.6 &      2.95 \\
 Ti\,I &   4534.780 &      0.84 &     0.336 &     39.3 &      2.20  &     88.8 &      3.02 \\
 Ti\,I &   4535.567 &      0.83 &     0.120 &   \nodata&   \nodata  &     81.5 &      3.10 \\
 Ti\,I &   4544.687 &      0.82 &  $-$0.520 &   \nodata&   \nodata  &     51.6 &      3.23 \\
 Ti\,I &   4656.470 &      0.00 &  $-$1.289 &   \nodata&   \nodata  &     62.6 &      3.17 \\
 Ti\,I &   4681.910 &      0.05 &  $-$1.015 &     18.3 &      2.13  &     85.8 &      3.32 \\
 Ti\,I &   4840.870 &      0.90 &  $-$0.453 &   \nodata&   \nodata  &     36.4 &      2.98 \\
 Ti\,I &   4991.070 &      0.84 &     0.436 &   \nodata&   \nodata  &    101.4 &      3.08 \\
 Ti\,I &   4999.500 &      0.83 &     0.306 &   \nodata&   \nodata  &     97.4 &      3.12 \\
 Ti\,I &   5007.206 &      0.82 &     0.168 &   \nodata&   \nodata  &     89.6 &      3.10 \\
 Ti\,I &   5020.024 &      0.84 &  $-$0.358 &   \nodata&   \nodata  &     62.5 &      3.22 \\
 Ti\,I &   5024.843 &      0.82 &  $-$0.546 &   \nodata&   \nodata  &     37.6 &      2.99 \\
 Ti\,I &   5035.902 &      1.46 &     0.260 &   \nodata&   \nodata  &     60.5 &      3.32 \\
 Ti\,I &   5036.463 &      1.44 &     0.186 &   \nodata&   \nodata  &     59.6 &      3.35 \\
 Ti\,I &   5038.396 &      1.43 &     0.069 &   \nodata&   \nodata  &     32.9 &      3.02 \\
 Ti\,I &   5039.960 &      0.02 &  $-$1.130 &   \nodata&   \nodata  &     69.9 &      3.10 \\
 Ti\,I &   5064.650 &      0.05 &  $-$0.935 &   \nodata&   \nodata  &     72.2 &      2.98 \\
 Ti\,I &   5173.740 &      0.00 &  $-$1.062 &     32.3 &      2.40  &     83.3 &      3.19 \\
 Ti\,I &   5192.970 &      0.02 &  $-$0.950 &   \nodata&   \nodata  &     82.9 &      3.10 \\
 Ti\,I &   5210.390 &      0.05 &  $-$0.828 &     36.6 &      2.32  &     91.0 &      3.15 \\
Ti\,II &   4012.396 &      0.57 &  $-$1.750 &     86.4 &      2.25  &    120.7 &      3.01 \\
Ti\,II &   4025.120 &      0.61 &  $-$1.980 &     79.5 &      2.38  &    106.2 &      2.97 \\
Ti\,II &   4028.338 &      1.89 &  $-$0.960 &     53.4 &      2.40  &    113.8 &      3.68 \\
Ti\,II &   4161.527 &      1.08 &  $-$2.160 &     30.6 &      2.19  &     92.3 &      3.39 \\
Ti\,II &   4163.634 &      2.59 &  $-$0.400 &   \nodata&   \nodata  &     79.0 &      3.19 \\
Ti\,II &   4290.219 &      1.16 &  $-$0.930 &     97.0 &      2.33  &    157.0 &      3.55 \\
Ti\,II &   4300.049 &      1.18 &  $-$0.490 &    103.6 &      2.07  &    178.7 &      3.49 \\
Ti\,II &   4330.723 &      1.18 &  $-$2.060 &   \nodata&   \nodata  &     79.0 &      3.13 \\
Ti\,II &   4337.914 &      1.08 &  $-$0.960 &     88.4 &      2.05  &    134.4 &      3.02 \\
Ti\,II &   4394.059 &      1.22 &  $-$1.780 &     46.0 &      2.24  &    111.9 &      3.53 \\
Ti\,II &   4395.031 &      1.08 &  $-$0.540 &    132.6 &      2.62  &    186.3 &      3.49 \\
Ti\,II &   4395.839 &      1.24 &  $-$1.930 &     37.0 &      2.25  &     79.2 &      3.07 \\
Ti\,II &   4399.765 &      1.24 &  $-$1.190 &     61.7 &      1.95  &    117.8 &      3.08 \\
Ti\,II &   4409.520 &      1.23 &  $-$2.370 &   \nodata&   \nodata  &     64.2 &      3.23 \\
Ti\,II &   4417.714 &      1.17 &  $-$1.190 &     91.8 &      2.46  &    147.5 &      3.59 \\
Ti\,II &   4418.331 &      1.24 &  $-$1.970 &     49.8 &      2.52  &     84.0 &      3.19 \\
Ti\,II &   4441.731 &      1.18 &  $-$2.410 &     23.5 &      2.38  &     63.9 &      3.21 \\
Ti\,II &   4443.801 &      1.08 &  $-$0.720 &    118.8 &      2.48  &    152.9 &      3.10 \\
Ti\,II &   4444.554 &      1.12 &  $-$2.240 &     30.3 &      2.29  &     71.8 &      3.09 \\
Ti\,II &   4450.482 &      1.08 &  $-$1.520 &     87.7 &      2.58  &    122.8 &      3.29 \\
Ti\,II &   4464.448 &      1.16 &  $-$1.810 &     47.4 &      2.22  &    100.8 &      3.24 \\
Ti\,II &   4468.517 &      1.13 &  $-$0.600 &    103.0 &      2.06  &    170.4 &      3.35 \\
Ti\,II &   4470.853 &      1.17 &  $-$2.020 &     41.9 &      2.34  &     84.4 &      3.15 \\
Ti\,II &   4493.522 &      1.08 &  $-$3.020 &   \nodata&   \nodata  &     45.1 &      3.39 \\
Ti\,II &   4501.270 &      1.12 &  $-$0.770 &    123.1 &      2.66  &    172.1 &      3.52 \\
Ti\,II &   4529.480 &      1.57 &  $-$2.030 &   \nodata&   \nodata  &     89.5 &      3.73 \\
Ti\,II &   4533.960 &      1.24 &  $-$0.530 &    116.4 &      2.41  &    149.1 &      3.01 \\
Ti\,II &   4545.144 &      1.13 &  $-$1.810 &     37.7 &      2.00  &     89.0 &      2.96 \\
Ti\,II &   4563.770 &      1.22 &  $-$0.960 &     99.3 &      2.41  &    157.4 &      3.56 \\
Ti\,II &   4571.971 &      1.57 &  $-$0.320 &     99.2 &      2.20  &    140.6 &      3.03 \\
Ti\,II &   4583.409 &      1.16 &  $-$2.920 &   \nodata&   \nodata  &     30.7 &      3.12 \\
Ti\,II &   4589.915 &      1.24 &  $-$1.790 &     63.1 &      2.55  &    116.9 &      3.61 \\
Ti\,II &   4657.200 &      1.24 &  $-$2.240 &   \nodata&   \nodata  &     88.8 &      3.50 \\
Ti\,II &   4708.662 &      1.24 &  $-$2.340 &   \nodata&   \nodata  &     78.1 &      3.41 \\
Ti\,II &   4779.979 &      2.05 &  $-$1.370 &   \nodata&   \nodata  &     72.5 &      3.32 \\
Ti\,II &   4798.532 &      1.08 &  $-$2.680 &     21.8 &      2.46  &     67.5 &      3.37 \\
Ti\,II &   4805.089 &      2.06 &  $-$1.100 &     47.8 &      2.57  &     92.5 &      3.41 \\
Ti\,II &   4865.610 &      1.12 &  $-$2.810 &   \nodata&   \nodata  &     69.8 &      3.58 \\
Ti\,II &   5129.156 &      1.89 &  $-$1.240 &     30.7 &      2.18  &     86.1 &      3.20 \\
Ti\,II &   5185.902 &      1.89 &  $-$1.490 &   \nodata&   \nodata  &     56.2 &      2.96 \\
Ti\,II &   5188.687 &      1.58 &  $-$1.050 &     60.8 &      2.13  &    142.5 &      3.64 \\
Ti\,II &   5226.538 &      1.57 &  $-$1.260 &     57.5 &      2.26  &    111.6 &      3.25 \\
Ti\,II &   5268.607 &      2.60 &  $-$1.620 &   \nodata&   \nodata  &     31.2 &      3.50 \\
Ti\,II &   5336.786 &      1.58 &  $-$1.590 &   \nodata&   \nodata  &     86.4 &      3.15 \\
Ti\,II &   5381.021 &      1.57 &  $-$1.920 &   \nodata&   \nodata  &     67.5 &      3.16 \\
Ti\,II &   5418.768 &      1.58 &  $-$2.000 &   \nodata&   \nodata  &     63.9 &      3.19 \\
 Cr\,I &   4254.332 &      0.00 &  $-$0.114 &     95.7 &      2.01  &    173.7 &      3.33 \\
 Cr\,I &   4274.800 &      0.00 &  $-$0.220 &     89.1 &      1.95  &    183.3 &      3.57 \\
 Cr\,I &   4289.720 &      0.00 &  $-$0.370 &     83.1 &      1.96  &    182.0 &      3.69 \\
 Cr\,I &   4337.566 &      0.97 &  $-$1.112 &   \nodata&   \nodata  &     68.9 &      3.50 \\
 Cr\,I &   4545.950 &      0.94 &  $-$1.370 &   \nodata&   \nodata  &     62.8 &      3.60 \\
 Cr\,I &   4580.056 &      0.94 &  $-$1.650 &   \nodata&   \nodata  &     58.6 &      3.81 \\
 Cr\,I &   4600.752 &      1.00 &  $-$1.260 &   \nodata&   \nodata  &     63.1 &      3.56 \\
 Cr\,I &   4616.137 &      0.98 &  $-$1.190 &   \nodata&   \nodata  &     60.5 &      3.42 \\
 Cr\,I &   4626.188 &      0.97 &  $-$1.320 &   \nodata&   \nodata  &     43.9 &      3.27 \\
 Cr\,I &   4646.150 &      1.03 &  $-$0.740 &   \nodata&   \nodata  &    102.7 &      3.78 \\
 Cr\,I &   4651.280 &      0.98 &  $-$1.460 &   \nodata&   \nodata  &     59.3 &      3.67 \\
 Cr\,I &   4652.158 &      1.00 &  $-$1.030 &   \nodata&   \nodata  &     71.1 &      3.46 \\
 Cr\,I &   5206.040 &      0.94 &     0.020 &     72.5 &      2.37  &    139.4 &      3.49 \\
 Cr\,I &   5208.419 &      0.94 &     0.160 &     76.7 &      2.31  &    175.3 &      3.93 \\
 Cr\,I &   5247.564 &      0.96 &  $-$1.640 &   \nodata&   \nodata  &     46.4 &      3.58 \\
 Cr\,I &   5296.690 &      0.98 &  $-$1.360 &   \nodata&   \nodata  &     65.7 &      3.62 \\
 Cr\,I &   5298.280 &      0.98 &  $-$1.140 &   \nodata&   \nodata  &     91.1 &      3.80 \\
 Cr\,I &   5345.800 &      1.00 &  $-$0.950 &   \nodata&   \nodata  &     77.7 &      3.41 \\
 Cr\,I &   5348.310 &      1.00 &  $-$1.210 &   \nodata&   \nodata  &     48.1 &      3.21 \\
 Cr\,I &   5409.770 &      1.03 &  $-$0.670 &     21.0 &      2.19  &    103.5 &      3.61 \\
Cr\,II &   4558.594 &      4.07 &  $-$0.656 &   \nodata&   \nodata  &     38.3 &      3.53 \\
Cr\,II &   4588.142 &      4.07 &  $-$0.826 &   \nodata&   \nodata  &     40.8 &      3.75 \\
 Mn\,I &   4030.753 &      HFS  & \nodata   &    107.5 &      2.00  &      syn &      2.80 \\
 Mn\,I &   4033.062 &      HFS  & \nodata   &   \nodata&   \nodata  &      syn &      2.80 \\
 Mn\,I &   4034.483 &      HFS  & \nodata   &     88.4 &      1.85  &      syn &      3.10 \\
 Mn\,I &   4041.357 &      HFS  & \nodata   &   \nodata&   \nodata  &      syn &      3.07 \\
 Mn\,I &   4754.048 &      HFS  & \nodata   &   \nodata&   \nodata  &      syn &      2.95 \\
 Mn\,I &   4783.432 &      HFS  & \nodata   &   \nodata&   \nodata  &      syn &      2.95 \\
 Mn\,I &   4823.528 &      HFS  & \nodata   &   \nodata&   \nodata  &      syn &      2.90 \\
 Fe\,I &   3753.611 &      2.18 &  $-$0.890 &     60.7 &      4.16  &   \nodata&   \nodata \\
 Fe\,I &   3765.539 &      3.24 &     0.482 &     71.3 &      4.29  &   \nodata&   \nodata \\
 Fe\,I &   3786.677 &      1.01 &  $-$2.185 &     86.2 &      4.66  &   \nodata&   \nodata \\
 Fe\,I &   3849.967 &      1.01 &  $-$0.863 &    116.0 &      4.08  &   \nodata&   \nodata \\
 Fe\,I &   3852.573 &      2.18 &  $-$1.180 &     71.5 &      4.68  &    110.7 &      5.48 \\
 Fe\,I &   3865.523 &      1.01 &  $-$0.950 &    128.5 &      4.44  &   \nodata&   \nodata \\
 Fe\,I &   3887.048 &      0.91 &  $-$1.140 &    116.4 &      4.23  &   \nodata&   \nodata \\
 Fe\,I &   3940.878 &      0.96 &  $-$2.600 &     67.9 &      4.52  &    125.4 &      5.67 \\
 Fe\,I &   3949.953 &      2.18 &  $-$1.251 &     53.5 &      4.35  &    121.2 &      5.75 \\
 Fe\,I &   3977.741 &      2.20 &  $-$1.120 &     59.2 &      4.35  &   \nodata&   \nodata \\
 Fe\,I &   4005.242 &      1.56 &  $-$0.583 &    110.0 &      4.25  &   \nodata&   \nodata \\
 Fe\,I &   4021.866 &      2.76 &  $-$0.730 &   \nodata&   \nodata  &     84.0 &      5.07 \\
 Fe\,I &   4032.627 &      1.49 &  $-$2.380 &     42.5 &      4.43  &    115.7 &      5.86 \\
 Fe\,I &   4044.609 &      2.83 &  $-$1.220 &   \nodata&   \nodata  &     77.1 &      5.50 \\
 Fe\,I &   4058.217 &      3.21 &  $-$1.110 &   \nodata&   \nodata  &     65.0 &      5.61 \\
 Fe\,I &   4062.441 &      2.85 &  $-$0.860 &   \nodata&   \nodata  &     76.2 &      5.14 \\
 Fe\,I &   4063.594 &      1.56 &     0.062 &    134.4 &      4.12  &   \nodata&   \nodata \\
 Fe\,I &   4067.978 &      3.21 &  $-$0.470 &     41.8 &      4.56  &     84.8 &      5.35 \\
 Fe\,I &   4070.769 &      3.24 &  $-$0.790 &   \nodata&   \nodata  &     74.9 &      5.50 \\
 Fe\,I &   4071.738 &      1.61 &  $-$0.008 &    138.8 &      4.32  &   \nodata&   \nodata \\
 Fe\,I &   4073.763 &      3.27 &  $-$0.900 &   \nodata&   \nodata  &     63.6 &      5.44 \\
 Fe\,I &   4076.629 &      3.21 &  $-$0.370 &   \nodata&   \nodata  &    105.8 &      5.70 \\
 Fe\,I &   4079.838 &      2.86 &  $-$1.360 &   \nodata&   \nodata  &     66.2 &      5.46 \\
 Fe\,I &   4095.970 &      2.59 &  $-$1.480 &   \nodata&   \nodata  &     77.5 &      5.47 \\
 Fe\,I &   4098.176 &      3.24 &  $-$0.880 &   \nodata&   \nodata  &     68.9 &      5.48 \\
 Fe\,I &   4109.802 &      2.85 &  $-$0.940 &   \nodata&   \nodata  &    102.4 &      5.77 \\
 Fe\,I &   4114.445 &      2.83 &  $-$1.303 &   \nodata&   \nodata  &     52.4 &      5.12 \\
 Fe\,I &   4132.058 &      1.61 &  $-$0.675 &    113.4 &      4.45  &   \nodata&   \nodata \\
 Fe\,I &   4132.899 &      2.85 &  $-$1.010 &     38.3 &      4.60  &     72.2 &      5.20 \\
 Fe\,I &   4134.678 &      2.83 &  $-$0.649 &     42.3 &      4.30  &     93.9 &      5.25 \\
 Fe\,I &   4136.998 &      3.42 &  $-$0.450 &     34.9 &      4.65  &     60.3 &      5.10 \\
 Fe\,I &   4139.927 &      0.99 &  $-$3.629 &   \nodata&   \nodata  &     88.3 &      5.87 \\
 Fe\,I &   4143.414 &      3.05 &  $-$0.200 &     42.4 &      4.11  &    102.7 &      5.26 \\
 Fe\,I &   4143.868 &      1.56 &  $-$0.511 &    130.6 &      4.59  &   \nodata&   \nodata \\
 Fe\,I &   4147.669 &      1.48 &  $-$2.071 &     63.4 &      4.50  &    131.0 &      5.82 \\
 Fe\,I &   4152.169 &      0.96 &  $-$3.232 &   \nodata&   \nodata  &     99.9 &      5.66 \\
 Fe\,I &   4153.899 &      3.40 &  $-$0.320 &   \nodata&   \nodata  &     89.4 &      5.50 \\
 Fe\,I &   4154.498 &      2.83 &  $-$0.688 &   \nodata&   \nodata  &     87.0 &      5.15 \\
 Fe\,I &   4154.805 &      3.37 &  $-$0.400 &     44.8 &      4.73  &     80.5 &      5.37 \\
 Fe\,I &   4156.799 &      2.83 &  $-$0.808 &   \nodata&   \nodata  &    104.9 &      5.66 \\
 Fe\,I &   4157.780 &      3.42 &  $-$0.403 &   \nodata&   \nodata  &     67.3 &      5.18 \\
 Fe\,I &   4158.793 &      3.43 &  $-$0.670 &   \nodata&   \nodata  &     71.2 &      5.53 \\
 Fe\,I &   4174.913 &      0.91 &  $-$2.938 &     63.7 &      4.67  &    129.1 &      5.92 \\
 Fe\,I &   4175.636 &      2.85 &  $-$0.827 &   \nodata&   \nodata  &     88.1 &      5.33 \\
 Fe\,I &   4181.755 &      2.83 &  $-$0.371 &     57.1 &      4.29  &    118.5 &      5.51 \\
 Fe\,I &   4182.382 &      3.02 &  $-$1.180 &     28.9 &      4.78  &     51.4 &      5.20 \\
 Fe\,I &   4184.892 &      2.83 &  $-$0.869 &   \nodata&   \nodata  &     83.8 &      5.26 \\
 Fe\,I &   4187.039 &      2.45 &  $-$0.514 &     86.1 &      4.58  &    124.3 &      5.26 \\
 Fe\,I &   4187.795 &      2.42 &  $-$0.510 &     69.4 &      4.17  &    118.8 &      5.11 \\
 Fe\,I &   4191.430 &      2.47 &  $-$0.666 &     62.9 &      4.27  &    111.2 &      5.18 \\
 Fe\,I &   4195.329 &      3.33 &  $-$0.492 &     40.9 &      4.70  &     74.8 &      5.30 \\
 Fe\,I &   4196.208 &      3.40 &  $-$0.700 &   \nodata&   \nodata  &     75.7 &      5.60 \\
 Fe\,I &   4199.095 &      3.05 &     0.156 &     79.9 &      4.51  &    116.9 &      5.21 \\
 Fe\,I &   4202.029 &      1.49 &  $-$0.689 &    118.6 &      4.41  &   \nodata&   \nodata \\
 Fe\,I &   4216.184 &      0.00 &  $-$3.357 &     93.1 &      4.63  &   \nodata&   \nodata \\
 Fe\,I &   4217.545 &      3.43 &  $-$0.484 &   \nodata&   \nodata  &     81.9 &      5.53 \\
 Fe\,I &   4222.213 &      2.45 &  $-$0.914 &     48.6 &      4.21  &    106.2 &      5.28 \\
 Fe\,I &   4227.427 &      3.33 &     0.266 &     65.4 &      4.41  &    124.6 &      5.58 \\
 Fe\,I &   4233.603 &      2.48 &  $-$0.579 &     61.0 &      4.15  &    115.0 &      5.17 \\
 Fe\,I &   4238.810 &      3.40 &  $-$0.233 &   \nodata&   \nodata  &     73.8 &      5.09 \\
 Fe\,I &   4247.426 &      3.37 &  $-$0.240 &     37.1 &      4.41  &     83.9 &      5.25 \\
 Fe\,I &   4250.119 &      2.47 &  $-$0.380 &     69.6 &      4.10  &    129.7 &      5.24 \\
 Fe\,I &   4250.787 &      1.56 &  $-$0.713 &    118.7 &      4.50  &   \nodata&   \nodata \\
 Fe\,I &   4260.474 &      2.40 &     0.077 &    106.8 &      4.39  &   \nodata&   \nodata \\
 Fe\,I &   4271.154 &      2.45 &  $-$0.337 &     76.0 &      4.17  &   \nodata&   \nodata \\
 Fe\,I &   4271.760 &      1.49 &  $-$0.173 &    143.5 &      4.36  &   \nodata&   \nodata \\
 Fe\,I &   4282.403 &      2.18 &  $-$0.779 &     66.1 &      4.09  &    123.7 &      5.20 \\
 Fe\,I &   4325.762 &      1.61 &     0.006 &    138.0 &      4.21  &   \nodata&   \nodata \\
 Fe\,I &   4337.046 &      1.56 &  $-$1.695 &     85.6 &      4.67  &    131.4 &      5.48 \\
 Fe\,I &   4352.735 &      2.22 &  $-$1.290 &     48.5 &      4.30  &    105.6 &      5.35 \\
 Fe\,I &   4375.930 &      0.00 &  $-$3.005 &    112.6 &      4.69  &   \nodata&   \nodata \\
 Fe\,I &   4388.407 &      3.60 &  $-$0.681 &   \nodata&   \nodata  &     55.8 &      5.44 \\
 Fe\,I &   4404.750 &      1.56 &  $-$0.147 &    144.9 &      4.41  &   \nodata&   \nodata \\
 Fe\,I &   4407.709 &      2.18 &  $-$1.970 &     25.2 &      4.47  &    100.0 &      5.85 \\
 Fe\,I &   4415.122 &      1.61 &  $-$0.621 &    110.4 &      4.23  &   \nodata&   \nodata \\
 Fe\,I &   4422.568 &      2.85 &  $-$1.110 &   \nodata&   \nodata  &     71.0 &      5.23 \\
 Fe\,I &   4427.310 &      0.05 &  $-$2.924 &     98.7 &      4.33  &   \nodata&   \nodata \\
 Fe\,I &   4430.614 &      2.22 &  $-$1.659 &     49.1 &      4.68  &    107.3 &      5.73 \\
 Fe\,I &   4442.339 &      2.20 &  $-$1.228 &     44.2 &      4.13  &    134.0 &      5.81 \\
 Fe\,I &   4443.194 &      2.86 &  $-$1.043 &     24.4 &      4.33  &     77.8 &      5.30 \\
 Fe\,I &   4447.717 &      2.22 &  $-$1.339 &     52.1 &      4.41  &    121.0 &      5.69 \\
 Fe\,I &   4454.381 &      2.83 &  $-$1.300 &   \nodata&   \nodata  &     65.3 &      5.30 \\
 Fe\,I &   4459.118 &      2.18 &  $-$1.279 &     42.7 &      4.13  &    137.6 &      5.90 \\
 Fe\,I &   4461.653 &      0.09 &  $-$3.194 &     89.7 &      4.42  &   \nodata&   \nodata \\
 Fe\,I &   4466.552 &      2.83 &  $-$0.600 &     47.0 &      4.30  &    120.2 &      5.69 \\
 Fe\,I &   4476.019 &      2.85 &  $-$0.820 &     45.9 &      4.53  &    109.5 &      5.70 \\
 Fe\,I &   4489.739 &      0.12 &  $-$3.899 &     39.0 &      4.18  &    120.0 &      5.57 \\
 Fe\,I &   4490.083 &      3.02 &  $-$1.580 &   \nodata&   \nodata  &     36.5 &      5.31 \\
 Fe\,I &   4494.563 &      2.20 &  $-$1.143 &     68.0 &      4.48  &    114.6 &      5.31 \\
 Fe\,I &   4528.614 &      2.18 &  $-$0.822 &     68.4 &      4.14  &    127.9 &      5.23 \\
 Fe\,I &   4531.148 &      1.48 &  $-$2.101 &     43.9 &      4.12  &    144.7 &      5.97 \\
 Fe\,I &   4592.651 &      1.56 &  $-$2.462 &     44.6 &      4.58  &     98.3 &      5.49 \\
 Fe\,I &   4595.359 &      3.29 &  $-$1.758 &   \nodata&   \nodata  &     22.9 &      5.53 \\
 Fe\,I &   4602.941 &      1.49 &  $-$2.208 &     51.4 &      4.37  &    123.9 &      5.66 \\
 Fe\,I &   4630.120 &      2.28 &  $-$2.587 &   \nodata&   \nodata  &     49.3 &      5.65 \\
 Fe\,I &   4632.912 &      1.61 &  $-$2.913 &   \nodata&   \nodata  &     76.5 &      5.59 \\
 Fe\,I &   4647.434 &      2.95 &  $-$1.351 &   \nodata&   \nodata  &     68.9 &      5.53 \\
 Fe\,I &   4678.846 &      3.60 &  $-$0.830 &   \nodata&   \nodata  &     40.3 &      5.31 \\
 Fe\,I &   4691.411 &      2.99 &  $-$1.520 &   \nodata&   \nodata  &     63.8 &      5.66 \\
 Fe\,I &   4707.274 &      3.24 &  $-$1.080 &   \nodata&   \nodata  &     73.5 &      5.68 \\
 Fe\,I &   4710.283 &      3.02 &  $-$1.610 &   \nodata&   \nodata  &     53.9 &      5.62 \\
 Fe\,I &   4733.591 &      1.49 &  $-$2.988 &   \nodata&   \nodata  &    101.2 &      5.96 \\
 Fe\,I &   4736.772 &      3.21 &  $-$0.752 &     29.7 &      4.56  &     83.9 &      5.50 \\
 Fe\,I &   4786.806 &      3.00 &  $-$1.606 &   \nodata&   \nodata  &     51.8 &      5.56 \\
 Fe\,I &   4789.650 &      3.53 &  $-$0.957 &   \nodata&   \nodata  &     39.1 &      5.33 \\
 Fe\,I &   4859.741 &      2.88 &  $-$0.760 &     56.3 &      4.66  &    104.2 &      5.48 \\
 Fe\,I &   4871.318 &      2.87 &  $-$0.362 &     61.4 &      4.34  &    117.4 &      5.31 \\
 Fe\,I &   4872.137 &      2.88 &  $-$0.567 &     38.0 &      4.14  &    104.9 &      5.30 \\
 Fe\,I &   4890.755 &      2.88 &  $-$0.394 &     48.3 &      4.15  &    128.9 &      5.56 \\
 Fe\,I &   4891.492 &      2.85 &  $-$0.111 &     79.6 &      4.41  &    131.6 &      5.29 \\
 Fe\,I &   4903.310 &      2.88 &  $-$0.926 &     29.  &      4.32  &     89.7 &      5.36 \\
 Fe\,I &   4918.994 &      2.85 &  $-$0.342 &     60.0 &      4.27  &    120.6 &      5.32 \\
 Fe\,I &   4920.503 &      2.83 &     0.068 &    105.2 &      4.74  &    136.1 &      5.16 \\
 Fe\,I &   4924.770 &      2.28 &  $-$2.114 &     25.8 &      4.71  &     59.0 &      5.30 \\
 Fe\,I &   4938.814 &      2.88 &  $-$1.077 &     40.2 &      4.69  &     81.2 &      5.36 \\
 Fe\,I &   4939.687 &      0.86 &  $-$3.252 &     39.8 &      4.41  &    117.2 &      5.69 \\
 Fe\,I &   4946.388 &      3.37 &  $-$1.170 &   \nodata&   \nodata  &     39.2 &      5.34 \\
 Fe\,I &   4966.089 &      3.33 &  $-$0.871 &   \nodata&   \nodata  &     71.7 &      5.53 \\
 Fe\,I &   4994.130 &      0.92 &  $-$2.969 &     41.4 &      4.23  &    121.5 &      5.55 \\
 Fe\,I &   5001.870 &      3.88 &     0.050 &   \nodata&   \nodata  &     88.5 &      5.57 \\
 Fe\,I &   5006.119 &      2.83 &  $-$0.615 &   \nodata&   \nodata  &    110.0 &      5.35 \\
 Fe\,I &   5012.068 &      0.86 &  $-$2.642 &     80.8 &      4.53  &    150.7 &      5.66 \\
 Fe\,I &   5014.942 &      3.94 &  $-$0.300 &   \nodata&   \nodata  &     48.7 &      5.30 \\
 Fe\,I &   5022.236 &      3.98 &  $-$0.530 &   \nodata&   \nodata  &     39.3 &      5.42 \\
 Fe\,I &   5041.072 &      0.96 &  $-$3.090 &     71.6 &      4.92  &    111.3 &      5.52 \\
 Fe\,I &   5041.756 &      1.49 &  $-$2.200 &   \nodata&   \nodata  &    106.1 &      5.21 \\
 Fe\,I &   5044.212 &      2.85 &  $-$2.017 &   \nodata&   \nodata  &     44.9 &      5.66 \\
 Fe\,I &   5049.820 &      2.28 &  $-$1.355 &     60.2 &      4.58  &    107.7 &      5.36 \\
 Fe\,I &   5051.634 &      0.92 &  $-$2.764 &     66.6 &      4.44  &   \nodata&   \nodata \\
 Fe\,I &   5074.749 &      4.22 &  $-$0.200 &   \nodata&   \nodata  &     52.6 &      5.60 \\
 Fe\,I &   5079.224 &      2.20 &  $-$2.105 &   \nodata&   \nodata  &     94.6 &      5.77 \\
 Fe\,I &   5079.740 &      0.99 &  $-$3.245 &   \nodata&   \nodata  &    112.6 &      5.73 \\
 Fe\,I &   5083.339 &      0.96 &  $-$2.842 &     47.8 &      4.25  &    136.5 &      5.72 \\
 Fe\,I &   5098.697 &      2.18 &  $-$2.030 &     32.8 &      4.65  &    100.1 &      5.78 \\
 Fe\,I &   5110.413 &      0.00 &  $-$3.760 &     72.1 &      4.40  &   \nodata&   \nodata \\
 Fe\,I &   5123.720 &      1.01 &  $-$3.058 &     58.4 &      4.71  &    116.6 &      5.64 \\
 Fe\,I &   5125.117 &      4.22 &  $-$0.140 &   \nodata&   \nodata  &     62.3 &      5.70 \\
 Fe\,I &   5127.360 &      0.92 &  $-$3.249 &     55.6 &      4.74  &    119.0 &      5.76 \\
 Fe\,I &   5131.468 &      2.22 &  $-$2.515 &   \nodata&   \nodata  &     62.5 &      5.67 \\
 Fe\,I &   5133.689 &      4.18 &     0.140 &   \nodata&   \nodata  &     79.2 &      5.66 \\
 Fe\,I &   5141.739 &      2.42 &  $-$2.238 &   \nodata&   \nodata  &     53.3 &      5.49 \\
 Fe\,I &   5142.929 &      0.96 &  $-$3.080 &     44.5 &      4.43  &    127.8 &      5.79 \\
 Fe\,I &   5150.840 &      0.99 &  $-$3.037 &     45.9 &      4.45  &    124.2 &      5.72 \\
 Fe\,I &   5151.911 &      1.01 &  $-$3.321 &   \nodata&   \nodata  &     97.5 &      5.55 \\
 Fe\,I &   5162.273 &      4.18 &     0.020 &   \nodata&   \nodata  &     56.6 &      5.39 \\
 Fe\,I &   5166.282 &      0.00 &  $-$4.123 &     67.6 &      4.68  &    122.5 &      5.48 \\
 Fe\,I &   5171.596 &      1.49 &  $-$1.721 &     72.3 &      4.20  &    149.2 &      5.48 \\
 Fe\,I &   5191.455 &      3.04 &  $-$0.551 &     41.0 &      4.35  &    105.1 &      5.43 \\
 Fe\,I &   5192.344 &      3.00 &  $-$0.421 &   \nodata&   \nodata  &    118.6 &      5.50 \\
 Fe\,I &   5194.942 &      1.56 &  $-$2.021 &     51.5 &      4.21  &    118.9 &      5.32 \\
 Fe\,I &   5198.711 &      2.22 &  $-$2.091 &   \nodata&   \nodata  &     80.0 &      5.52 \\
 Fe\,I &   5202.336 &      2.18 &  $-$1.871 &   \nodata&   \nodata  &    105.3 &      5.70 \\
 Fe\,I &   5216.274 &      1.61 &  $-$2.082 &     55.7 &      4.41  &    128.7 &      5.62 \\
 Fe\,I &   5217.390 &      3.21 &  $-$1.162 &   \nodata&   \nodata  &     76.4 &      5.73 \\
 Fe\,I &   5225.526 &      0.11 &  $-$4.755 &   \nodata&   \nodata  &     98.5 &      5.84 \\
 Fe\,I &   5232.940 &      2.94 &  $-$0.057 &     72.1 &      4.29  &    124.1 &      5.16 \\
 Fe\,I &   5242.491 &      3.63 &  $-$0.967 &   \nodata&   \nodata  &     37.2 &      5.40 \\
 Fe\,I &   5247.050 &      0.09 &  $-$4.946 &   \nodata&   \nodata  &     80.1 &      5.72 \\
 Fe\,I &   5250.210 &      0.12 &  $-$4.938 &   \nodata&   \nodata  &     82.8 &      5.79 \\
 Fe\,I &   5250.646 &      2.20 &  $-$2.180 &     27.0 &      4.69  &     94.2 &      5.82 \\
 Fe\,I &   5254.956 &      0.11 &  $-$4.764 &     19.8 &      4.56  &    101.0 &      5.89 \\
 Fe\,I &   5263.305 &      3.27 &  $-$0.879 &   \nodata&   \nodata  &     93.0 &      5.81 \\
 Fe\,I &   5266.555 &      3.00 &  $-$0.385 &     57.5 &      4.43  &    113.8 &      5.37 \\
 Fe\,I &   5269.537 &      0.86 &  $-$1.333 &    146.1 &      4.54  &   \nodata&   \nodata \\
 Fe\,I &   5281.790 &      3.04 &  $-$0.833 &   \nodata&   \nodata  &     98.5 &      5.57 \\
 Fe\,I &   5283.621 &      3.24 &  $-$0.524 &     31.4 &      4.37  &    107.6 &      5.68 \\
 Fe\,I &   5302.300 &      3.28 &  $-$0.720 &   \nodata&   \nodata  &     81.6 &      5.46 \\
 Fe\,I &   5307.361 &      1.61 &  $-$2.912 &   \nodata&   \nodata  &     82.6 &      5.62 \\
 Fe\,I &   5322.040 &      2.28 &  $-$2.802 &   \nodata&   \nodata  &     34.6 &      5.56 \\
 Fe\,I &   5324.179 &      3.21 &  $-$0.103 &     62.5 &      4.47  &    114.5 &      5.34 \\
 Fe\,I &   5328.039 &      0.92 &  $-$1.466 &    147.2 &      4.76  &   \nodata&   \nodata \\
 Fe\,I &   5328.531 &      1.56 &  $-$1.850 &     78.1 &      4.51  &    134.4 &      5.41 \\
 Fe\,I &   5332.900 &      1.55 &  $-$2.776 &   \nodata&   \nodata  &    109.7 &      5.87 \\
 Fe\,I &   5339.930 &      3.27 &  $-$0.720 &   \nodata&   \nodata  &     98.4 &      5.73 \\
 Fe\,I &   5364.871 &      4.45 &     0.228 &   \nodata&   \nodata  &     57.9 &      5.51 \\
 Fe\,I &   5365.400 &      3.56 &  $-$1.020 &   \nodata&   \nodata  &     27.7 &      5.17 \\
 Fe\,I &   5367.467 &      4.42 &     0.443 &   \nodata&   \nodata  &     71.3 &      5.49 \\
 Fe\,I &   5369.962 &      4.37 &     0.536 &   \nodata&   \nodata  &     87.8 &      5.62 \\
 Fe\,I &   5371.489 &      0.96 &  $-$1.644 &    127.5 &      4.60  &   \nodata&   \nodata \\
 Fe\,I &   5379.573 &      3.69 &  $-$1.514 &   \nodata&   \nodata  &     25.3 &      5.76 \\
 Fe\,I &   5383.369 &      4.31 &     0.645 &   \nodata&   \nodata  &     73.7 &      5.19 \\
 Fe\,I &   5389.479 &      4.42 &  $-$0.410 &   \nodata&   \nodata  &     24.4 &      5.50 \\
 Fe\,I &   5393.168 &      3.24 &  $-$0.910 &   \nodata&   \nodata  &     91.7 &      5.77 \\
 Fe\,I &   5397.128 &      0.92 &  $-$1.982 &    126.7 &      4.87  &   \nodata&   \nodata \\
 Fe\,I &   5405.775 &      0.99 &  $-$1.852 &    126.9 &      4.82  &   \nodata&   \nodata \\
 Fe\,I &   5410.910 &      4.47 &     0.398 &   \nodata&   \nodata  &     50.5 &      5.24 \\
 Fe\,I &   5415.199 &      4.39 &     0.643 &     25.5 &      4.44  &     65.4 &      5.16 \\
 Fe\,I &   5424.068 &      4.32 &     0.520 &   \nodata&   \nodata  &     76.6 &      5.38 \\
 Fe\,I &   5429.696 &      0.96 &  $-$1.881 &    120.1 &      4.67  &   \nodata&   \nodata \\
 Fe\,I &   5434.524 &      1.01 &  $-$2.126 &     83.8 &      4.21  &    168.4 &      5.51 \\
 Fe\,I &   5446.917 &      0.99 &  $-$1.910 &     94.2 &      4.18  &   \nodata&   \nodata \\
 Fe\,I &   5455.609 &      1.01 &  $-$2.090 &    118.7 &      4.91  &   \nodata&   \nodata \\
 Fe\,I &   5497.516 &      1.01 &  $-$2.825 &     69.4 &      4.63  &    127.0 &      5.52 \\
 Fe\,I &   5501.465 &      0.96 &  $-$3.046 &     49.8 &      4.46  &    120.3 &      5.55 \\
 Fe\,I &   5506.779 &      0.99 &  $-$2.789 &     83.0 &      4.82  &    130.5 &      5.51 \\
 Fe\,I &   5569.618 &      3.42 &  $-$0.540 &   \nodata&   \nodata  &     82.2 &      5.43 \\
 Fe\,I &   5572.842 &      3.40 &  $-$0.275 &   \nodata&   \nodata  &     95.0 &      5.37 \\
 Fe\,I &   5586.756 &      3.37 &  $-$0.144 &     33.9 &      4.18  &    123.2 &      5.69 \\
 Fe\,I &   5615.644 &      3.33 &     0.050 &   \nodata&   \nodata  &    133.7 &      5.62 \\
 Fe\,I &   5624.542 &      3.42 &  $-$0.755 &   \nodata&   \nodata  &     74.1 &      5.52 \\
 Fe\,I &   5658.816 &      3.40 &  $-$0.793 &   \nodata&   \nodata  &     82.8 &      5.66 \\
 Fe\,I &   5662.516 &      4.18 &  $-$0.573 &   \nodata&   \nodata  &     35.8 &      5.61 \\
 Fe\,I &   5686.529 &      4.55 &  $-$0.450 &   \nodata&   \nodata  &     15.4 &      5.44 \\
 Fe\,I &   5701.544 &      2.56 &  $-$2.143 &   \nodata&   \nodata  &     46.4 &      5.41 \\
 Fe\,I &   6065.481 &      2.61 &  $-$1.410 &   \nodata&   \nodata  &     97.7 &      5.53 \\
 Fe\,I &   6136.615 &      2.45 &  $-$1.410 &     30.2 &      4.24  &     94.1 &      5.27 \\
 Fe\,I &   6137.691 &      2.59 &  $-$1.346 &     48.3 &      4.68  &    107.2 &      5.61 \\
 Fe\,I &   6151.618 &      2.18 &  $-$3.371 &   \nodata&   \nodata  &     22.6 &      5.72 \\
 Fe\,I &   6191.558 &      2.43 &  $-$1.416 &     39.7 &      4.41  &    121.5 &      5.72 \\
 Fe\,I &   6200.312 &      2.61 &  $-$2.437 &   \nodata&   \nodata  &     38.5 &      5.62 \\
 Fe\,I &   6213.429 &      2.22 &  $-$2.481 &   \nodata&   \nodata  &     63.1 &      5.57 \\
 Fe\,I &   6219.280 &      2.20 &  $-$2.448 &   \nodata&   \nodata  &     67.9 &      5.58 \\
 Fe\,I &   6230.723 &      2.56 &  $-$1.276 &     42.3 &      4.47  &    114.2 &      5.61 \\
 Fe\,I &   6240.645 &      2.22 &  $-$3.173 &   \nodata&   \nodata  &     27.1 &      5.66 \\
 Fe\,I &   6246.318 &      3.60 &  $-$0.877 &   \nodata&   \nodata  &     48.2 &      5.41 \\
 Fe\,I &   6252.555 &      2.40 &  $-$1.687 &   \nodata&   \nodata  &    105.3 &      5.66 \\
 Fe\,I &   6254.257 &      2.28 &  $-$2.443 &   \nodata&   \nodata  &     70.4 &      5.70 \\
 Fe\,I &   6265.134 &      2.18 &  $-$2.540 &   \nodata&   \nodata  &     65.2 &      5.61 \\
 Fe\,I &   6297.792 &      2.22 &  $-$2.640 &   \nodata&   \nodata  &     41.5 &      5.39 \\
 Fe\,I &   6301.499 &      3.65 &  $-$0.718 &   \nodata&   \nodata  &     64.4 &      5.57 \\
 Fe\,I &   6322.684 &      2.59 &  $-$2.469 &   \nodata&   \nodata  &     42.2 &      5.68 \\
 Fe\,I &   6335.330 &      2.20 &  $-$2.180 &   \nodata&   \nodata  &     70.0 &      5.34 \\
 Fe\,I &   6336.835 &      3.69 &  $-$1.050 &   \nodata&   \nodata  &     50.8 &      5.72 \\
 Fe\,I &   6344.148 &      2.43 &  $-$2.877 &   \nodata&   \nodata  &     26.9 &      5.61 \\
 Fe\,I &   6393.601 &      2.43 &  $-$1.576 &     25.0 &      4.26  &    114.7 &      5.73 \\
 Fe\,I &   6400.000 &      3.60 &  $-$0.290 &     24.4 &      4.37  &     69.0 &      5.14 \\
 Fe\,I &   6411.649 &      3.65 &  $-$0.595 &   \nodata&   \nodata  &     60.6 &      5.37 \\
 Fe\,I &   6421.350 &      2.28 &  $-$2.014 &     23.6 &      4.48  &    102.2 &      5.76 \\
 Fe\,I &   6430.846 &      2.18 &  $-$1.946 &     39.7 &      4.62  &    115.1 &      5.80 \\
 Fe\,I &   6494.980 &      2.40 &  $-$1.239 &     52.6 &      4.39  &    126.7 &      5.55 \\
 Fe\,I &   6498.940 &      0.96 &  $-$4.699 &   \nodata&   \nodata  &     31.9 &      5.73 \\
 Fe\,I &   6592.912 &      2.73 &  $-$1.473 &   \nodata&   \nodata  &     91.2 &      5.60 \\
 Fe\,I &   6593.868 &      2.44 &  $-$2.366 &   \nodata&   \nodata  &     57.4 &      5.62 \\
 Fe\,I &   6609.109 &      2.56 &  $-$2.661 &   \nodata&   \nodata  &     34.0 &      5.68 \\
 Fe\,I &   6663.440 &      2.42 &  $-$2.479 &   \nodata&   \nodata  &     43.1 &      5.48 \\
 Fe\,I &   6677.986 &      2.69 &  $-$1.418 &     27.8 &      4.47  &     94.6 &      5.54 \\
 Fe\,I &   6750.152 &      2.42 &  $-$2.584 &   \nodata&   \nodata  &     50.5 &      5.69 \\
 Fe\,I &   6978.850 &      2.48 &  $-$2.452 &   \nodata&   \nodata  &     54.9 &      5.69 \\
Fe\,II &   4122.668 &      2.58 &  $-$3.380 &   \nodata&   \nodata  &     60.9 &      5.75 \\
Fe\,II &   4233.170 &      2.58 &  $-$1.970 &     76.1 &      4.59  &    116.2 &      5.46 \\
Fe\,II &   4416.817 &      2.78 &  $-$2.600 &     38.2 &      4.72  &     66.1 &      5.27 \\
Fe\,II &   4489.185 &      2.83 &  $-$2.970 &   \nodata&   \nodata  &     61.3 &      5.62 \\
Fe\,II &   4491.410 &      2.86 &  $-$2.710 &   \nodata&   \nodata  &     51.1 &      5.21 \\
Fe\,II &   4508.283 &      2.86 &  $-$2.580 &     20.1 &      4.39  &     81.7 &      5.63 \\
Fe\,II &   4515.340 &      2.84 &  $-$2.600 &     31.3 &      4.64  &     79.0 &      5.58 \\
Fe\,II &   4520.224 &      2.81 &  $-$2.600 &   \nodata&   \nodata  &     60.6 &      5.21 \\
Fe\,II &   4522.630 &      2.84 &  $-$2.250 &     50.8 &      4.66  &    106.2 &      5.77 \\
Fe\,II &   4541.523 &      2.86 &  $-$3.050 &   \nodata&   \nodata  &     62.2 &      5.74 \\
Fe\,II &   4555.890 &      2.83 &  $-$2.400 &     37.4 &      4.55  &     85.6 &      5.49 \\
Fe\,II &   4576.340 &      2.84 &  $-$2.950 &   \nodata&   \nodata  &     51.2 &      5.43 \\
Fe\,II &   4583.840 &      2.81 &  $-$1.930 &     47.2 &      4.23  &    106.8 &      5.42 \\
Fe\,II &   4620.520 &      2.83 &  $-$3.210 &   \nodata&   \nodata  &     27.4 &      5.23 \\
Fe\,II &   4731.439 &      2.89 &  $-$3.360 &   \nodata&   \nodata  &     44.3 &      5.77 \\
Fe\,II &   4923.930 &      2.89 &  $-$1.320 &     76.8 &      4.24  &    134.3 &      5.39 \\
Fe\,II &   4993.350 &      2.81 &  $-$3.670 &   \nodata&   \nodata  &     20.9 &      5.50 \\
Fe\,II &   5018.450 &      2.89 &  $-$1.220 &     83.2 &      4.25  &    156.7 &      5.69 \\
Fe\,II &   5197.580 &      3.23 &  $-$2.220 &     21.3 &      4.46  &     78.1 &      5.59 \\
Fe\,II &   5234.630 &      3.22 &  $-$2.180 &   \nodata&   \nodata  &     80.6 &      5.58 \\
Fe\,II &   5276.000 &      3.20 &  $-$2.010 &     24.9 &      4.30  &     93.8 &      5.62 \\
Fe\,II &   5284.080 &      2.89 &  $-$3.190 &   \nodata&   \nodata  &     33.9 &      5.39 \\
Fe\,II &   5534.834 &      3.25 &  $-$2.930 &   \nodata&   \nodata  &     40.5 &      5.67 \\
Fe\,II &   6247.545 &      3.89 &  $-$2.510 &   \nodata&   \nodata  &     25.7 &      5.68 \\
Fe\,II &   6432.680 &      2.89 &  $-$3.710 &   \nodata&   \nodata  &     20.0 &      5.56 \\
Fe\,II &   6456.383 &      3.90 &  $-$2.080 &   \nodata&   \nodata  &     34.4 &      5.44 \\
 Co\,I &   3845.468 &      0.92 &     0.010 &     98.1 &      2.50  &   \nodata&   \nodata \\
 Co\,I &   3873.120 &      0.43 &  $-$0.660 &     98.2 &      2.57  &   \nodata&   \nodata \\
 Co\,I &   3995.306 &      0.92 &  $-$0.220 &     84.5 &      2.33  &    117.0 &      2.89 \\
 Co\,I &   4110.532 &      1.05 &  $-$1.080 &   \nodata&   \nodata  &     64.8 &      2.82 \\
 Co\,I &   4118.767 &      1.05 &  $-$0.490 &   \nodata&   \nodata  &    109.8 &      3.12 \\
 Co\,I &   4121.318 &      0.92 &  $-$0.320 &     85.4 &      2.41  &    104.9 &      2.68 \\
 Ni\,I &   3423.700 &      0.21 &  $-$0.790 &   \nodata&   \nodata  &    162.3 &      3.71 \\
 Ni\,I &   3483.770 &      0.28 &  $-$1.120 &    107.5 &      3.23  &   \nodata&   \nodata \\
 Ni\,I &   3524.540 &      0.03 &     0.007 &    194.0 &      3.02  &   \nodata&   \nodata \\
 Ni\,I &   3597.710 &      0.21 &  $-$1.115 &   \nodata&   \nodata  &    394.2 &      5.25 \\
 Ni\,I &   3783.520 &      0.42 &  $-$1.420 &    101.1 &      3.30  &   \nodata&   \nodata \\
 Ni\,I &   3858.301 &      0.42 &  $-$0.951 &    135.8 &      3.58  &    125.3 &      3.12 \\
 Ni\,I &   4648.659 &      3.42 &  $-$0.160 &   \nodata&   \nodata  &     32.6 &      3.99 \\
 Ni\,I &   4714.421 &      3.38 &     0.230 &   \nodata&   \nodata  &     66.2 &      4.14 \\
 Ni\,I &   4980.161 &      3.61 &  $-$0.110 &   \nodata&   \nodata  &     53.2 &      4.51 \\
 Ni\,I &   5035.374 &      3.63 &     0.290 &   \nodata&   \nodata  &     45.3 &      4.00 \\
 Ni\,I &   5080.523 &      3.65 &     0.130 &   \nodata&   \nodata  &     35.1 &      4.00 \\
 Ni\,I &   5084.080 &      3.68 &     0.030 &   \nodata&   \nodata  &     25.9 &      3.95 \\
 Ni\,I &   5137.075 &      1.68 &  $-$1.990 &   \nodata&   \nodata  &     88.6 &      4.63 \\
 Ni\,I &   5155.760 &      3.90 &  $-$0.090 &   \nodata&   \nodata  &     23.1 &      4.26 \\
 Ni\,I &   5476.900 &      1.83 &  $-$0.890 &     46.8 &      3.03  &    120.4 &      4.25 \\
 Ni\,I &   6108.121 &      1.68 &  $-$2.450 &   \nodata&   \nodata  &     25.4 &      3.97 \\
 Ni\,I &   6643.640 &      1.68 &  $-$2.300 &   \nodata&   \nodata  &     64.0 &      4.43 \\
 Ni\,I &   6767.770 &      1.83 &  $-$2.170 &   \nodata&   \nodata  &     39.3 &      4.10 \\
 Zn\,I &   4810.528 &      4.08 &  $-$0.137 &     15.9 &      1.90  &  $<$40.0 &   $<$2.52 \\
Sr\,II &   4077.714 &      0.00 &     0.150 &    101.1 &   $-$1.09  &  \nodata &   \nodata \\
Sr\,II &   4215.524 &      0.00 &  $-$0.180 &    114.5 &   $-$1.19  &     syn  &   $-$0.32 \\
Ba\,II &   4554.033 &      HFS  &  \nodata  &     73.3 &   $-$2.08  &     syn  &   $-$0.51 \\
Ba\,II &   4934.086 &      HFS  &  \nodata  &     58.3 &   $-$2.13  &     syn  &   $-$0.61 \\
Ba\,II &   6496.896 &      HFS  &  \nodata  &   \nodata&   \nodata  &     syn  &   $-$0.48 \\
\enddata
\tablecomments{The table is available in its entirety only in
  electronic format. A portion is shown for guidance regarding its
  form and content.}
\end{deluxetable*}

Using 1D plane-parallel model atmospheres with $\alpha$-enhancement
from \citet{castelli_kurucz} and the latest version of the MOOG
analysis code \citep{moog, sobeck11}, we computed local thermodynamic
equilibrium (LTE) abundances for our Bo\"otes\,I stars. Final
abundance ratios [X/Fe] are obtained using the solar abundances of
\citet{asplund09} and listed in Table~\ref{abund}. Abundance
uncertainties refer to the standard deviation of the line abundance
measurements for each element. In case of elements with few lines and
the standard deviation resulting in very small values, we adopt a
nominal minimum uncertainty of 0.1\,dex.

\begin{deluxetable}{lrrrrrrrrrrrrrrrrrrrrrrrrrrrr}
\tabletypesize{\tiny}
\tablewidth{0pc}
\tablecaption{\label{abund} Magellan/MIKE Chemical Abundances of our two Bo\"otes\,I Stars}
\tablehead{
\colhead{Species} & 
\colhead{$N$} &
\colhead{$\log\epsilon (\mbox{X})$} & \colhead{$\sigma$}&  \colhead{[X/H]}& \colhead{[X/Fe]}} 
\startdata
\multicolumn{5}{c}{Boo-980}\\\hline
  CH    &   2 &    4.97 &    0.20 & $-$3.46 & $-$0.40  \\
  Na\,I &   2 &    3.51 &    0.10 & $-$2.73 &    0.33  \\
  Mg\,I &   7 &    4.83 &    0.13 & $-$2.77 &    0.29  \\ 
  Al\,I &   2 &    2.62 &    0.20 & $-$3.83 & $-$0.77  \\
  Si\,I &   2 &    4.88 &    0.10 & $-$2.62 &    0.44  \\ 
  Ca\,I &   8 &    3.57 &    0.11 & $-$2.77 &    0.30  \\
 Sc\,II &   6 &    0.08 &    0.10 & $-$3.07 & $-$0.01  \\
  Ti\,I &   8 &    2.28 &    0.15 & $-$2.67 &    0.40  \\
 Ti\,II &  31 &    2.32 &    0.19 & $-$2.63 &    0.43  \\
  Cr\,I &   6 &    2.13 &    0.17 & $-$3.51 & $-$0.45  \\
  Mn\,I &   3 &    1.93 &    0.10 & $-$3.50 & $-$0.44  \\
  Fe\,I & 125 &    4.44 &    0.20 & $-$3.06 &    0.00  \\
 Fe\,II &  11 &    4.46 &    0.18 & $-$3.04 &    0.02  \\
  Co\,I &   4 &    2.45 &    0.10 & $-$2.54 &    0.53  \\
  Ni\,I &   5 &    3.23 &    0.21 & $-$2.99 &    0.07  \\
  Zn\,I &   1 &    1.90 &    0.15 & $-$2.66 &    0.40  \\
 Sr\,II &   2 & $-$1.14 &    0.20 & $-$4.01 & $-$0.95  \\
 Ba\,II &   2 & $-$2.11 &    0.20 & $-$4.29 & $-$1.23  \\
 Eu\,II &   1 &$<-$2.44 & \nodata &$<-$2.96 & $<$0.10 \\\hline
\multicolumn{5}{c}{Boo-127}\\\hline
  CH    &   2 &    6.29 &    0.20 & $-$2.14 & $-$0.15  \\
  Na\,I &   2 &    4.23 &    0.10 & $-$2.01 & $-$0.02  \\
  Mg\,I &   6 &    6.02 &    0.11 & $-$1.58 &    0.41  \\ 
  Al\,I &   1 &    3.57 &    0.20 & $-$2.88 & $-$0.89  \\
  Si\,I &   1 &    6.28 &    0.20 & $-$1.23 &    0.76  \\ 
  Ca\,I &  21 &    4.60 &    0.13 & $-$1.74 &    0.24  \\
 Sc\,II &   9 &    1.07 &    0.10 & $-$2.08 & $-$0.09  \\ 
  Ti\,I &  22 &    3.11 &    0.12 & $-$1.84 &    0.15  \\
 Ti\,II &  46 &    3.30 &    0.22 & $-$1.65 &    0.33  \\
  Cr\,I &  20 &    3.57 &    0.18 & $-$2.07 & $-$0.09  \\
 Cr\,II &   2 &    3.64 &    0.11 & $-$2.00 & $-$0.01  \\
  Mn\,I &   7 &    2.94 &    0.14 & $-$2.49 & $-$0.50  \\
  Fe\,I & 204 &    5.51 &    0.21 & $-$1.99 &    0.00  \\
 Fe\,II &  26 &    5.53 &    0.17 & $-$1.97 &    0.01  \\
  Co\,I &   4 &    2.88 &    0.16 & $-$2.11 & $-$0.12  \\
  Ni\,I &  16 &    4.17 &    0.44 & $-$2.05 & $-$0.07  \\
  Zn\,I &   1 & $<$2.52 & \nodata &$<-$2.04 &$<-$0.05  \\       
 Sr\,II &   1 & $-$0.32 &    0.99 & $-$3.19 & $-$1.20\tablenotemark{a}  \\
 Ba\,II &   3 & $-$0.53 &    0.10 & $-$2.71 & $-$0.72  \\
 Eu\,II &   1 &$<-$1.37 & \nodata &$<-$1.89 & $<$0.10  
\enddata 
\tablenotetext{a}{Derived only from the $\lambda4215$
    line. The $\lambda4077$ line gave a spuriously low abundance.}
\end{deluxetable}

\subsection{Stellar parameters}
We derive stellar parameters spectroscopically from Fe\,I and Fe\,II
lines by minimizing abundance trends and matching the Fe\,I to Fe\,II
abundance. We follow the procedure described in \citet{frebel13} which
adjusts the spectroscopic temperatures to better match those derived
from photometry. In this way, no unphysically cool temperatures and
low surface gravities are obtained. We note that the present analysis
procedures are identical to those employed in the analysis of the
stars in Segue\,1 \citep{frebel14}. This allows us to make a
homogeneous comparison between the Bo\"otes\,I and Segue\,1 stars.
The final spectroscopic stellar parameters are T$_{\rm eff}=4720$\,K,
$\log g=1.4$, $v_{micr}=2.2$\,km\,s$^{-1}$ and $\mbox{[Fe/H]}=-3.06$
for Boo\,I-980 and T$_{\rm eff}=4765$\,K, $\log g=1.35$,
$v_{micr}=2.3$\,km\,s$^{-1}$ and $\mbox{[Fe/H]}=-1.99$ for Boo\,I-127.
We estimate our temperature uncertainties to be $\sim100$ - $150$\,K,
and those in $\log g$ and \mbox{v$_{\rm micr}$} to be 0.3\,dex and
0.3\,km\,s$^{-1}$, respectively. The spectroscopic temperatures agree
well with temperatures derived from the $(g-r)_o$ color transformed to
$B-V$ \citep{jordi06}, and using color-temperature relations from
\citet{alonso_giants}. They are T$_{\rm eff}=4810$\,K for Boo\,I-980
and T$_{\rm eff}=4625$\,K for Boo\,I-127. Our final spectroscopic
stellar parameters agree very well with values of a 12\,Gyr isochrone
with $\mbox{[$\alpha$/Fe]}=0.4$ and metallicities of $\mbox{[Fe/H]} =
-3.0$ and $-2.0$ \citep{Y2_iso}.



\subsection{Chemical abundances}


\begin{figure}[!ht]
\begin{center}
\includegraphics[clip=true, width=8.5cm, bbllx=54, bblly=110,
  bburx=440, bbury=768]{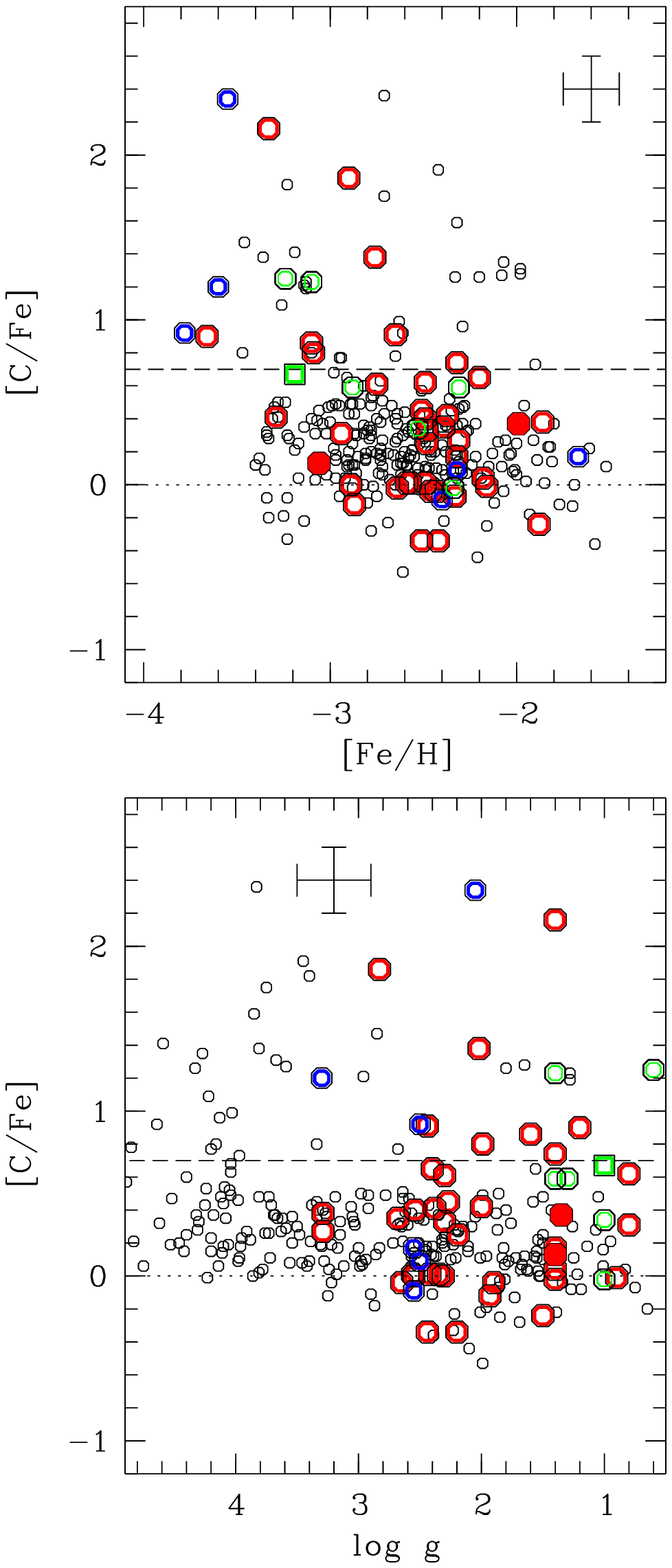}
\figcaption{\scriptsize \label{cfe} [C/Fe] abundance ratios (corrected
  for the evolutionary status of the stars) as a function of [Fe/H]
  (top panel) and surface gravity (bottom panel) of our Bo\"otes\,I
  stars (\textit{filled red circles}) and literature Bo\"otes\,I stars
  (\textit{open red circles}; \citealt{ishigaki14} values were used
  for the best possible internal consistency), in comparison with
  metal-poor halo stars (black circles; \citealt{heresII}) and other
  dwarf galaxy stars (Segue\,1: blue circles \citealt{frebel14};
  UMa\,II, ComBer and Leo\,IV: green circles
  \citealt{ufs,leo4}). Representative error bars are also shown.  The
  definition of C-enhancement from \citet{aoki_cemp_2007} is shown
  with a dashed line.}
\end{center}
\end{figure}

We here briefly summarize key aspects of our chemical abundance
analysis. Again, we refer the reader to \citet{frebel14} for further
details since our analysis follows the same procedures. We use the
4313\,{\AA} and 4323\,{\AA} CH features to determine the C abundance.
Both Boo-980 and Boo-127 have slightly subsolar [C/Fe] ratios, when
taken at face value.  However, one has to take into account that
carbon is converted to nitrogen due to the CN cycle operating on the
red giant branch.  Many of the Bo\"otes\,I stars are indeed located on
the giant branch and suffer from this depletion. Calculating
individual corrections based on stellar evolutionary modeling (see
\citealt{placco14} for more details) for each star to counter this
effect leads to final carbon abundances of $\mbox{[C/Fe]}=0.13$ and
$0.37$, for the two stars, respectively.  We also apply
  corrections to the adopted literature carbon abundances which range
  from 0\,dex for the warmer giants to 0.75\,dex for giants with
  T$_{\rm eff}=4500$\,K (see also Section~\ref{sec:comp} for further
discussion).  The corrected carbon abundances of all stars in
Bo\"otes\,I together with corrected values of the comparison stars are
shown in Figure~\ref{cfe}.

As can be seen in the figure, there is a clear trend of increasing
carbon abundance with decreasing [Fe/H], very similar to what is found
among halo stars (e.g., \citealt{heresII, placco14}). Overall, a range
in carbon abundances of 2.7\,dex is present among Bo\"otes\,I stars,
from $-0.34 < \mbox{[C/Fe]}<2.31$ (after applying the carbon
correction).  In fact, there are eight carbon-enhanced metal-poor
stars $\mbox{[C/Fe]}>0.7$ (hereafter ``CEMP stars''; CEMP
subclasses are discussed further below) after the application of the
carbon corrections of \citet{placco14}. (There are five CEMP stars
before the correction.) This means that the overall fraction of CEMP
stars eight out of 39, and thus $\sim21\%$. For stars with
$\mbox{[Fe/H]}<-2.5$, the fraction is seven out of 19, and thus
37\%. This increases to 66\% (four out of six) for
$\mbox{[Fe/H]}<-3.0$, and 100\% (one star) for
$\mbox{[Fe/H]}<-3.5$. These fractions of CEMP stars are considerably
higher than the results from halo stars for the respective
metallicities ranges, 24\%, 43\%, and 60\% (Placco et al. 2014). Note,
though, that the Bo\"otes\,I results are of course based on
significantly smaller samples. Nevertheless, we broadly conclude that
Bo\"otes\,I does not show a carbon abundance distribution that is
different from that of the Galactic halo.

Although also plagued by low number statistics, the same (qualitative)
behavior is found for the combined population of Segue\,1 (excluding
their metal-rich CH star which is not shown in Figure~\ref{cfe}), Ursa
Major\,II, Coma Berenices, and Leo\,IV stars. The overall fraction is
5 out 13 (38\%). No stars with $\mbox{[Fe/H]}>-3.0$ are CEMP stars but
the five stars with $\mbox{[Fe/H]}<-3.0$ are all carbon enhanced
(100\%).

\begin{figure*}[!th]
 \begin{center}
  \includegraphics[clip=true,width=18cm,bbllx=35, bblly=195, bburx=570,
   bbury=687]{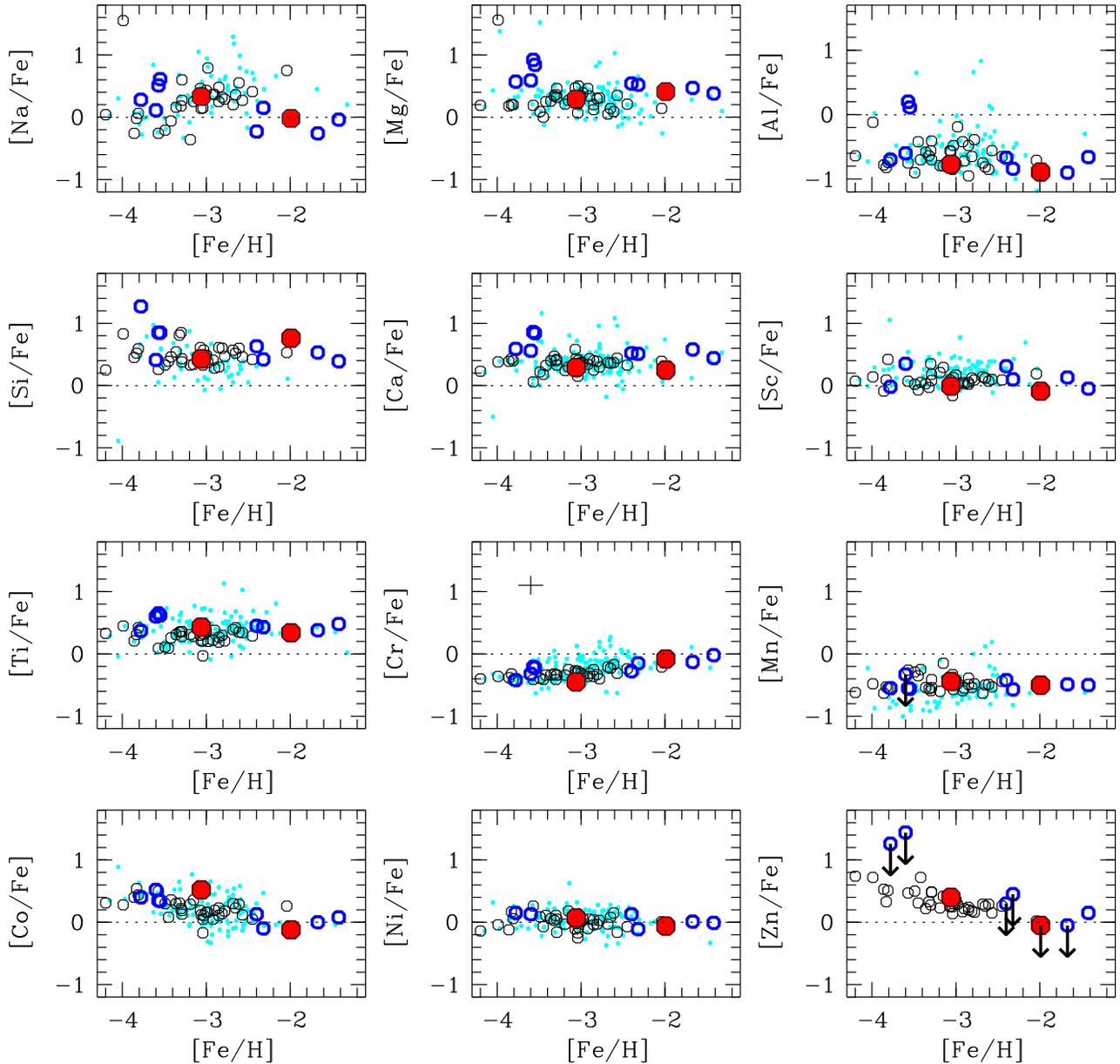} \figcaption{\small
     \label{cayrel_abundances} Abundance ratios ([X/Fe]) as a
     function of metallicity ([Fe/H]) for various elements detected in
     our Bo\"otes\,I stars (filled red circles) in comparison with those
     of stars in Segue\,1 \citep{frebel14,norris10_seg} (blue open
     circles) and of halo stars (black and small cyan circles) of
     \citet{cayrel2004} and \citet{yong13_II}, respectively. }
 \end{center}
\end{figure*}

Interestingly, two of the eight carbon enhanced stars in Bo\"otes\,I
for which barium abundances are available belong to the class of
CEMP-no stars, that is, CEMP stars without supersolar enhancement in
neutron-capture elements, i.e. $\mbox{[Ba/Fe]}<0$
\citep{araa}. Additional data would be needed to determine whether the
remaining six objects are also CEMP-no stars. It should be noted,
though, that no star with a barium measurement in Bo\"otes\,I has
enhanced neutron-capture element abundances compared to the solar
level, making it likely that these six objects will turn out to be
CEMP-no stars also. This speculation is furthermore supported by the
fact that all CEMP stars in the group of ultra-faint dwarfs of
Segue\,1 (SDSS\,J100714+160154, SDSS\,J100652+160235,
SDSS\,J100639+160008, Segue\,1-7), Ursa Major\,II (UMa\,II-S1), and
Leo\,IV (Leo\,IV-S1) stars are CEMP-no stars. For completeness, we
also note that one CEMP-s star was found in Segue~\,1. Interestingly,
several CEMP-r stars were recently found in the ultra-faint dwarf
galaxy Reticulum\,II (DES\,J033523−540407, DES J033607−540235, DES
J033454−540558) whose r-process elements likely stem from a
independent source than all other elements, namely a neutron star
merger or magnetar \citep{ji16b}.

Again, this behavior is very similar to that of halo stars which adds
evidence that the most metal-poor halo stars could have originated in
small dwarf galaxies such as the systems discussed here (see e.g.,
discussion in \citealt{fn15}). A further extensive discussion on the
carbon abundances in Bo\"otes\,I in the context of chemical evolution
can be found in \citet{gilmore13} and will not be repeated here.


We used the Mg\,I, Si\,I, Ca\,I and Ti\,II lines to
determine $\alpha$-element abundances from equivalent width
measurements and spectrum synthesis (in the case of Si). Our two stars
have enhanced, halo-typical $\alpha$-enhancement.
Figure~\ref{cayrel_abundances} shows the comparison of most of our
elemental abundances for the two stars, in comparison with those of
the extremely metal-poor star samples from \citet{cayrel2004},
\citet{yong13_II}, as well as the Segue\,1 ultra-faint dwarf galaxy
stars \citep{frebel14}. Overall, the agreement of the different groups
of stars is remarkably good.

\begin{figure}[!ht]
 \begin{center}
  \includegraphics[clip=true,width=6.5cm,bbllx=200, bblly=77, bburx=400,
   bbury=682]{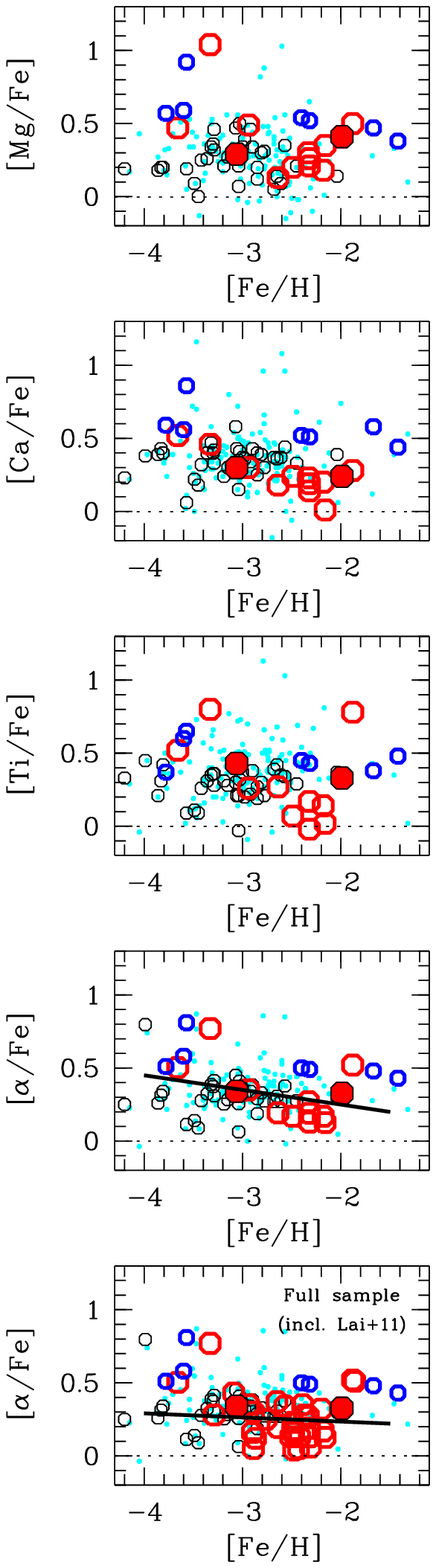} \figcaption{\scriptsize
     \label{abund_lit} Abundance ratios ([X/Fe]) of 
$\alpha$-elements of our two Bo\"otes\,I stars (full red circles)
     together with 11 Bo\"otes\,I stars (filled red circles) as analyzed by
     \citet{gilmore13}, \citet{ishigaki14}, and
     \citet{feltzing09}. Segue\,1 stellar abundances \citep{frebel14}
     are shown with blue circles, and halo stars in black circles
     \citep{cayrel2004} and cyan dots \citep{yong13_II}. The bottom
     two panels show combined [$\alpha$/Fe] ratios. The bottom panel
     additionally includes the [$\alpha$/Fe] of \citet{lai11}. Fits to
     the Bo\"otes data points (with Boo-119 excluded; [$\alpha$/Fe] =
     0.77) are also shown.}
 \end{center}
\end{figure}

Considering the full Bo\"otes\,I sample, a similar behavior is
found. About two-thirds of the Bo\"otes\,I stars with high-resolution
abundance measurements have halo-typical $\alpha$-element abundances,
as can be seen in Figure~\ref{abund_lit} in the top three panels. A
combined [$\alpha$/Fe] vs [Fe/H] (bottom two panels) then delivers an
overall reasonable agreement with halo star abundances. No obvious
downturn to solar-level ratios is found for Mg, Ca, Ti and the
combined [$\alpha$/Fe], despite some individual stars having low
abundances. Overall, though, the slope of [$\alpha$/Fe] does somewhat
decrease with increasing [Fe/H]. 

Excluding Boo-119, a CEMP-no star with $\mbox{[Fe/H]} =-3.3$
and [$\alpha$/Fe] = 0.77, the [$\alpha$/Fe] slope is $-$0.10 (rms
scatter of 0.12\,dex), based on the high-resolution abundances (second
lowest panel in the Figure). Also excluding Boo-41 (with
$\mbox{[Fe/H]} \sim -1.9$), as \citet{gilmore13} did, increases the
slope to $-$0.18. Gilmore et al. find $-0.19$, for comparison. When
adding the combined [$\alpha$/Fe] measurements of \citet{lai11}, the
slope decreases to $-$0.03 (bottom panel of Figure~\ref{abund_lit};
excluding only Boo-119), with an rms scatter of 0.13\,dex.
Implications of this behavior will be further discussed in
Section~\ref{signature}.

Iron-peak element abundances were obtained from lines as
listed in Table~\ref{Tab:Eqw}. The abundances found for our two
Bo\"otes\,I stars are also in excellent agreement with those of
comparable halo stars (see Figure~\ref{cayrel_abundances}). This
indicates the robust production of these elements in Bo\"otes\,I, in
the same way as in other dwarf galaxies and in the halo. This behavior
thus appears to be independent of environment.

\begin{figure}[!tb]
  \begin{center}
   \includegraphics[clip=true,width=9cm,bbllx=56, bblly=223,
     bburx=535, bbury=608]{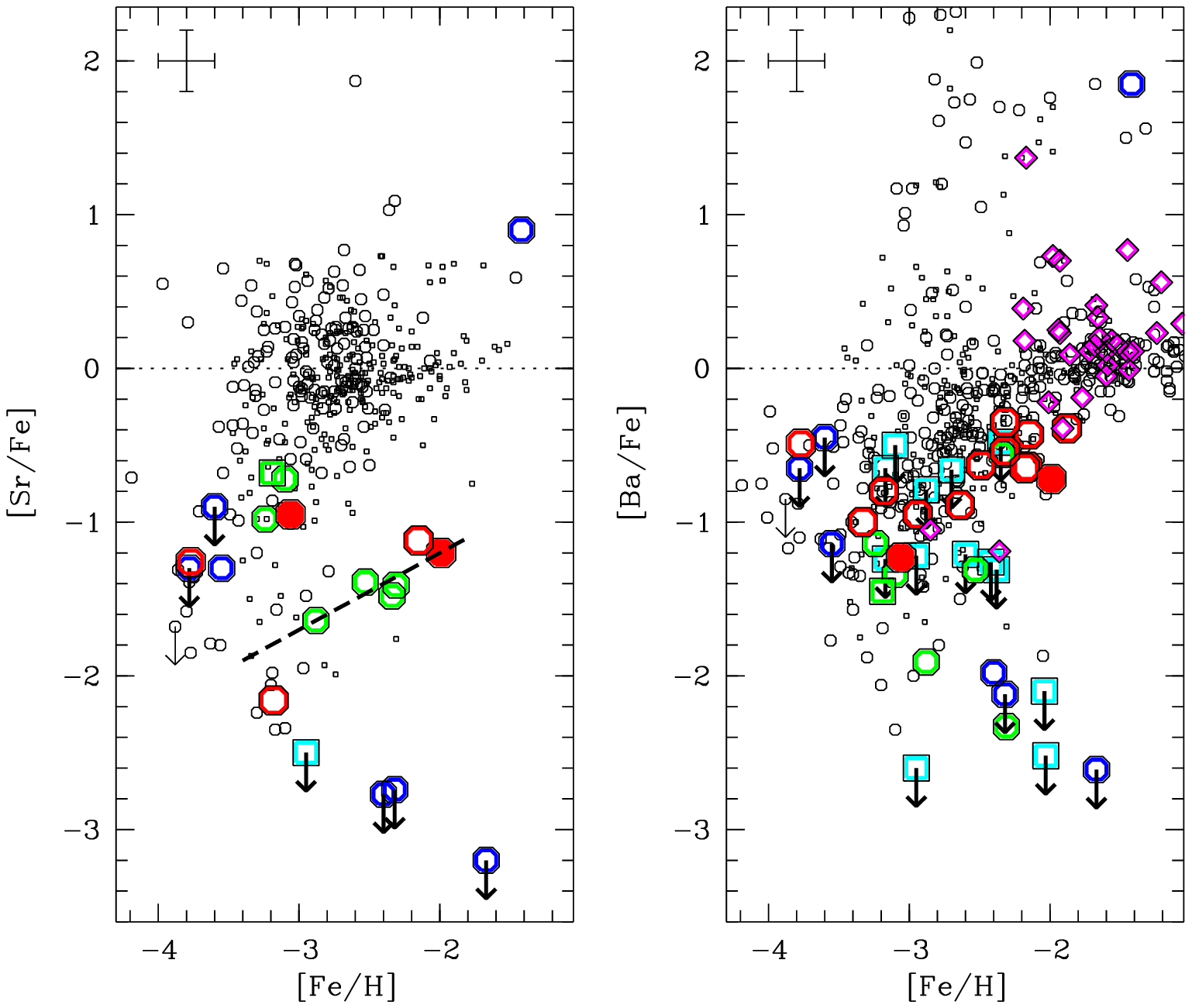}
   \includegraphics[clip=true,width=9cm,bbllx=56, bblly=223,
     bburx=535, bbury=608]{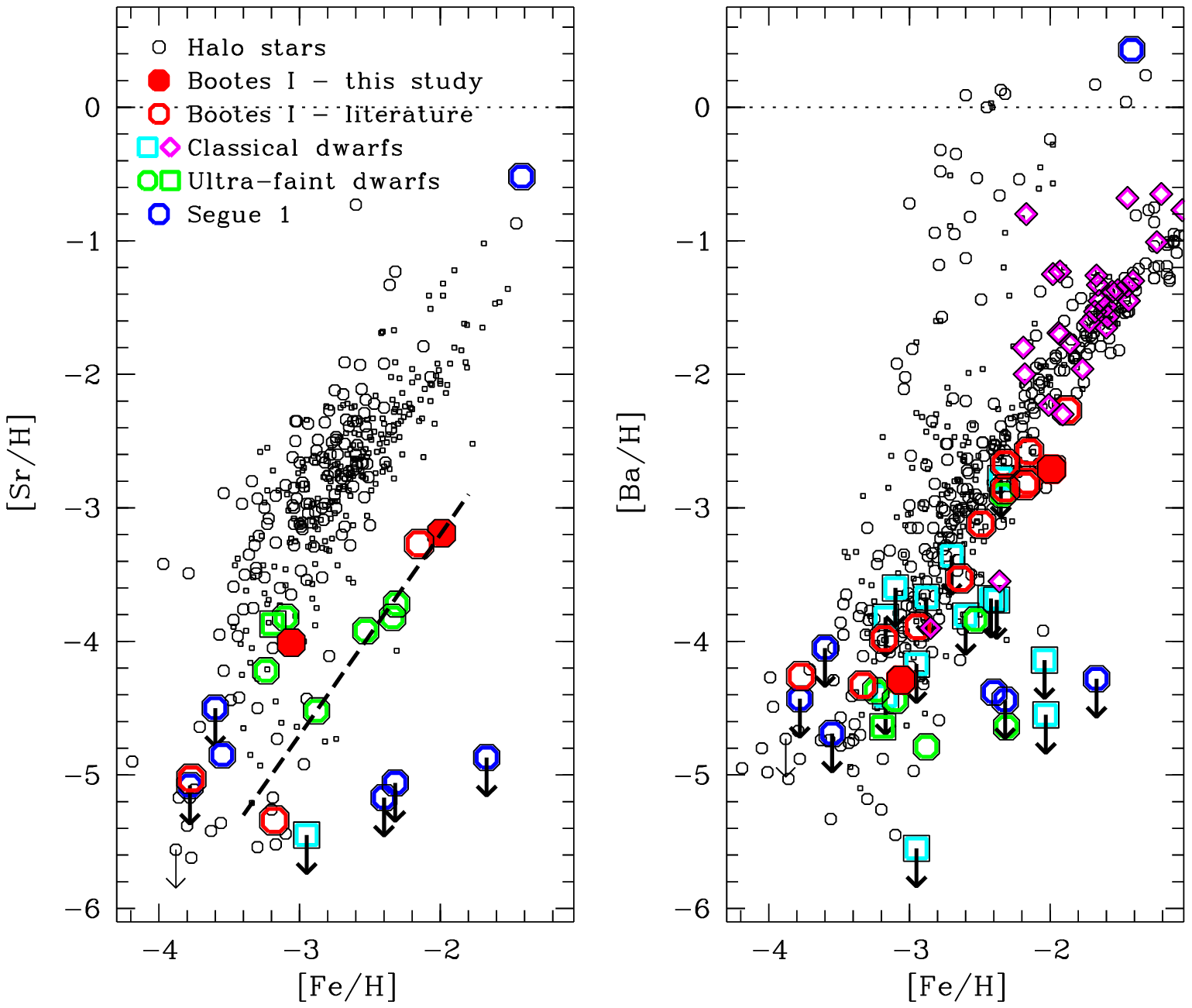}
   \figcaption{\scriptsize \label{ncap_plot} Abundance ratios of
     neutron-capture-elements [Sr/Fe] and [Ba/Fe] (top panels) and
     [Sr/H] and [Ba/H] (bottom panels) as a function of metallicity
     [Fe/H] of our Bo\"otes\,I stars (\textit{filled red circles}) and
     literature Bo\"otes\,I stars (\textit{open red circles}; here
     \citealt{ishigaki14} was used for Sr and \citealt{gilmore13} for
     Ba, for internal consistency) in comparison with those of other
     ultra faint dwarf galaxy stars in Segue\,1 (\textit{blue
       circles}; \citealt{frebel14,norris10_seg}), UMa\,II, ComBer,
     Leo\,IV (\textit{green circles}; \citealt{ufs, leo4}), Draco and
     Hercules (\textit{cyan squares}; \citealt{fulbright_rich,
       koch_her, koch13}), stars in the classical dwarf spheroidal
     galaxies (\textit{pink triangles}; \citealt{venn04}), and the
     galactic halo (\textit{black open circles}; \citealt{aoki05},
     \citealt{francois07}, \citealt{heresII} and
     \citealt{yong13_II}). The dashed line is meant to guide the eye
     regarding a putative second dwarf galaxy star sequence. Note that
     both axes have the same scale, showing the huge range of
     neutron-capture abundances in metal-poor stars. Representative
     error bars are shown in the [Sr/Fe] and [Ba/Fe] panels. }
  \end{center}
 \end{figure}

Finally, Sr and Ba are the only neutron-capture elements detectable in
the two stars. Boo-980 has Sr and Ba abundances similar to those of
halo stars with $\mbox{[Fe/H]} \sim -3$, although they are on the
lower end of that range. They are actually rather similar to the
abundances of Leo\,IV-SI in Leo\,IV studied by \citet{leo4} and two of
the Ursa Major\,II stars (UMa\,II-S1, UMa\,II-S2) of
\citet{frebel10}. This can be seen in Figure~\ref{ncap_plot}, where we
show both [Sr,Ba/Fe] and [Sr,Ba/H] to best illustrate the behavior of
neutron-capture elements.  Boo-127 shows a somewhat different
behavior. Its Sr abundance is about 1\,dex below the general Sr trend
set by halo stars. However, based on about half a dozen data points,
Boo-127 seems to extend an emerging second branch (marked in the
figure) of Sr abundance about one dex below the main halo trend. It
currently consists mainly of four Ursa Major\,II and Coma Berenices
stars, two more Bo\"otes\,I stars, together with a few halo
stars. Additional measurements of Sr in stars with $\mbox{[Fe/H]} >-2$
in ultra-faint dwarf galaxies would help to further investigate the
reality of this putative branch. Especially for Bo\"otes\,I, currently
only five stars have Sr measurements.

Boo-1137 also has a Sr abundance indistinguishable from halo
stars. However, its [Sr/H] level is extremely low at $\mbox{[Sr/H]}
\sim -5$. Boo-94 has an even lower value of $\mbox{[Sr/H]} \sim
-5.4$. Compared to Boo-127, this indicates a [Sr/H] spread of at least
2\,dex. The low level of Sr abundances in these two stars is
comparable with that found in Segue\,1. Generally, the Ba abundances
of all Bo\"otes\,I stars are somewhat uniformly offset from the main
halo star trend. The Ba abundance spread is $1.5$\,dex, and as in the
case of Sr, the lowest Ba abundances are similar to those found in
Segue\,1. This behavior will be discussed further in
Section~\ref{signature}. Interestingly, six Bo\"otes\,I stars
including Boo-127 cluster around $\mbox{[Ba/H]} \sim -2.7$ and
$\mbox{[Fe/H]} \sim-2.2$. These stars also have other similar
abundances and could perhaps represent a dissolved star cluster
candidate \citep{blandhawthorn10b,karlsson12} although more accurate
abundances are needed to draw firm conclusions.

\subsection{Comparison of Boo-127 abundances}\label{sec:comp}

Boo-127 was first observed by \citet{feltzing09} who found the star to
have unusually high Mg ($\mbox{[Mg/Fe]}=0.76$) and low Ca
($\mbox{[Ca/Fe]}=0.02$) abundances. However, these results were not
confirmed by \citet{gilmore13} or \citet{ishigaki14} who also included
this star in their samples.  Our independent observation of Boo-127 also yields
much more moderate values of $\mbox{[Mg/Fe]}=0.41$ and
$\mbox{[Ca/Fe]}=0.24$, in line with the Gilmore et al. and Ishigaki et
al. studies.  We thus conclude that Boo-127 has no unusual
$\alpha$-abundances, but rather a halo-like signature in these
elements.

\begin{figure}[!t]
 \begin{center}
  \includegraphics[clip=true,width=8cm,bbllx=200, bblly=565, bburx=500,
   bbury=767]{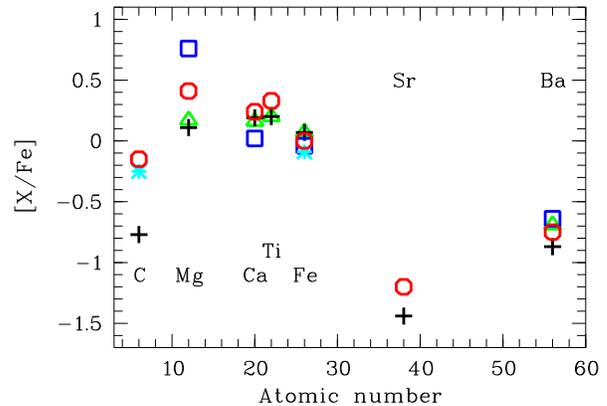} \figcaption{\small
     \label{boo_comp} Abundance ratios ([X/Fe]) of the four different studies that 
analyzed Boo-127 as a function of atomic number. In the case of iron,
[Fe/H] $-$ [Fe/H]$_{\rm{Boo-127,\,this\,study}}$ is shown on the y-axis.
Symbols are as follows: red circles: this work; green triangles:
\citet{gilmore13}; black crosses: \citet{ishigaki14}; blue squares:
\citet{feltzing09}; and cyan asterisk: \citet{norris10_booseg}.  }
 \end{center}
\end{figure}

Figure~\ref{boo_comp} shows a detailed comparison of various elements
observed in Boo-127 that are common to these four studies. The [C/Fe]
abundance of Norris et al. (2010b) is also included for
comparison. Except for the high Mg abundance of \citet{feltzing09} and
the low carbon abundance of \citet{ishigaki14}, the abundances agree
reasonably well and are within $\sim0.3$\,dex of each other. We note
here that \citet{ishigaki14} have systematically lower carbon
abundances compared to our study, \citet{norris10_booseg} and
\citet{lai11} by 0.43\,dex (based on five stars).  We note here that
the stellar parameters of all four studies are consistent, though. The
effective temperatures agree within 165\,K and the surface gravities
within 0.6\,dex. The [Fe/H] abundances are remarkably close and within
0.11\,dex. Excluding the \citet{feltzing09} results decreases these
ranges to 65\,K, 0.25\,dex and 0.07\,dex. Since the Mg\,b lines are
gravity sensitive (e.g., \citealt{hollek11}), the lower Feltzing et
al. gravity choice of $\log g=1.0$ might, in retrospect, explain their
high Mg abundance.

\subsection{Literature data on Bo\"otes\,I}

In Table~\ref{lit_abund} we list abundances of key elements common to
all studies of Bo\"otes\,I stars. We list each star's name as provided
by the original authors. We note that this results in an apparently
inconsistent system where, e.g., two distinct stars having the names
Boo-7 and Boo07. In reverse, where applicable, two distinct names for
the same star are also indicated. Coordinates for these stars can be
found in the references given in the last column.

For many stars, duplicate studies exist. Given authors who use
different data and analysis methods, to obtain a set of abundances as
internally consistent as possible between studies, and following the abundance
comparison in Figure~\ref{boo_comp}, we adopt a combination of
(adjusted) abundances for our analysis and in the figures
shown. Adjustments were made to reflect different iron abundances,
e.g., $\mbox{[C/Fe]}_{\rm adopted} = \mbox{[C/Fe]}_{\rm orig} +
\Delta\mbox{Fe/H]}$.

Generally, we adopted our own and the \citet{gilmore13} abundances (we
adopted an average of the ``NY'' and ``GM'' abundances) together with
the carbon values from \citet{norris10_booseg}. We adjusted the
[Sr/Fe] values from \citet{ishigaki14} to the Gilmore et al. scale. We
replaced the carbon abundance upper limits of \citet{ishigaki14} with
detections of \citet{norris10_booseg}. We used the adjusted carbon
abundance of \citet{norris10_booseg} together with the adopted
abundances of \citet{feltzing09} in the case of Boo-7.  Explanations
of which elemental abundances have been used are given in the table
notes but for clarity we list our adopted abundances for each star at
the bottom of the table.

Regarding abundance uncertainties in [X/Fe], systematic uncertainties
taking into account uncertainties in the stellar parameters as well as
random uncertainties are around 0.2 to 0.3\,dex for this kind of data
quality (e.g., Table~4 in \citealt{frebel14}). Keeping this in mind,
the various abundance averages are not significantly affected by our
attempt to construct a homogeneous abundance set and vary by on order
0.1\,dex. Nevertheless, even such subtle systematic changes could
affect the interpretation, and more generally, are important for
comparisons with chemical evolution models.

\begin{deluxetable*}{lrrrrrrrrrrll}
\tabletypesize{\tiny}
\tablewidth{0pc}
\tablecaption{\label{lit_abund} Chemical Abundances of Key Elements of
  Bo\"otes\,I Stars from the Literature}
\tablehead{
\colhead{Name} &
\colhead{\mbox{T$_{\rm eff}$}} & 
\colhead{$\log g$} &
\colhead{[Fe/H]}&  
\colhead{[C/Fe]}& 
\colhead{[Mg/Fe]}& 
\colhead{[Ca/Fe]}& 
\colhead{[Ti/Fe]}& 
\colhead{[Sr/Fe]}& 
\colhead{[Ba/Fe]}& 
\colhead{Comment}&
\colhead{Ref.}}
\startdata
 Boo02 &  5114 &  2.00 &$-$2.37 &  0.36  & 0.28\tablenotemark{a}&\nodata&\nodata&\nodata&  \nodata&med-res & LAI11\\
 Boo03 &  5127 &  1.99 &$-$3.09 &  0.79  & 0.43\tablenotemark{a}&\nodata&\nodata&\nodata&  \nodata&med-res & LAI11\\
 Boo04 &  5210 &  2.68 &$-$2.39 &   0.34 & 0.35\tablenotemark{a}&\nodata&\nodata&\nodata&  \nodata&med-res & LAI11\\
 Boo05 &  5077 &  2.55 &$-$2.89 &$<$0.00 & 0.05\tablenotemark{a}&\nodata&\nodata&\nodata&  \nodata&med-res & LAI11\\ 
 Boo06 &  5404 &  2.40 &$-$2.20 &   0.64 & 0.32\tablenotemark{a}&\nodata&\nodata&\nodata&  \nodata&med-res & LAI11\\
 Boo07 &  5200 &  2.54 &$-$2.49 &   0.39 & 0.16\tablenotemark{a}&\nodata&\nodata&\nodata&  \nodata&med-res & LAI11\\ 
 Boo-7 &  4800 &  1.60 &$-$2.32 &$-$0.50 &\nodata               &\nodata&\nodata&\nodata&  \nodata&med-res & NOR10\\
       &\nodata&\nodata&$-$2.33 & \nodata&  0.30                &  0.23 &\nodata&\nodata& $-$0.75 &high-res& FEL09\tablenotemark{b,c}\\
 Boo08 &  5178 &  2.30 &$-$2.48 &   0.32 & 0.18\tablenotemark{a}&\nodata&\nodata&\nodata&  \nodata&med-res & LAI11\\
 Boo-8 &  5090 &  2.30 &$-$2.75 &   0.60 &\nodata               &\nodata&\nodata&\nodata&  \nodata&med-res & NOR10\\
 Boo09 &  5563 &  2.44 &$-$2.65 &  0.90  & 0.37\tablenotemark{a}&\nodata&\nodata&\nodata&  \nodata&med-res & LAI11\\
 Boo-9 &  4630 &  1.10 &$-$2.67 &$-$0.55 &\nodata               &\nodata&\nodata&\nodata&  \nodata&med-res & NOR10\\
       &  4750 &  1.40 &$-$2.64 &$<-$0.29& 0.13                 &  0.18 & 0.27  &\nodata&$-$0.89  &high-res& ISH14\tablenotemark{b,c}\\
 Boo10 &  5086 &  2.42 &$-$2.59 &$<$0.00 &\nodata               &\nodata&\nodata&\nodata&  \nodata&med-res & LAI11\\
 Boo11 &  5199 &  2.65 &$-$2.43 &$-$0.05 & 0.05\tablenotemark{a}&\nodata&\nodata&\nodata&  \nodata&med-res & LAI11\\
 Boo12 &  5168 &  2.19 &$-$2.48 &  0.24  & 0.04\tablenotemark{a}&\nodata&\nodata&\nodata&  \nodata&med-res & LAI11\\
 Boo13 &  5631 &  2.35 &$-$2.49 &$<$0.00 & 0.23\tablenotemark{a}&\nodata&\nodata&\nodata&  \nodata&med-res & LAI11\\
 Boo14 &  5971 &  2.56 &$-$2.57 &$<$0.00 & 0.34\tablenotemark{a}&\nodata&\nodata&\nodata&  \nodata&med-res & LAI11\\
 Boo15 &  5117 &  2.31 &$-$2.89 &$<$0.00 & 0.30\tablenotemark{a}&\nodata&\nodata&\nodata&  \nodata&med-res & LAI11\\
 Boo18 &  5287 &  2.27 &$-$2.51 &   0.44 & 0.13\tablenotemark{a}&\nodata&\nodata&\nodata&  \nodata&med-res & LAI11\\
 Boo19 &  5141 &  2.38 &$-$3.29 &   0.40 & 0.28\tablenotemark{a}&\nodata&\nodata&\nodata&  \nodata&med-res & LAI11\\
 Boo20 &  4931 &  2.44 &$-$2.42 &$-$0.35 & 0.13\tablenotemark{a}&\nodata&\nodata&\nodata&  \nodata&med-res & LAI11\\
 Boo22 &  4866 &  1.93 &$-$2.87 &$-$0.16 & 0.12\tablenotemark{a}&\nodata&\nodata&\nodata&  \nodata&med-res & LAI11\\
 Boo23 &  5475 &  2.83 &$-$2.90 &  1.86  & 0.16\tablenotemark{a}&\nodata&\nodata&\nodata&  \nodata&med-res & LAI11\\
 Boo25 &  5141 &  2.02 &$-$2.76 &   1.34 & 0.25\tablenotemark{a}&\nodata&\nodata&\nodata&  \nodata&med-res & LAI11\\
 Boo28 &  5449 &  3.29 &$-$2.31 &   0.27 & 0.06\tablenotemark{a}&\nodata&\nodata&\nodata&  \nodata&med-res & LAI11\\
 Boo30 &  5449 &  3.29 &$-$1.86 &  0.38  & 0.51\tablenotemark{a}&\nodata&\nodata&\nodata&  \nodata&med-res & LAI11\\
 Boo-33&  4730 &  1.40 &$-$2.29 &  0.30  &\nodata               &\nodata&\nodata&\nodata&  \nodata&med-res & NOR10\\
       &  4600 &  1.00 &$-$2.52 &\nodata & 0.69                 &  0.40 &\nodata&\nodata&$-$0.40  &high-res& FEL09\\
       &  4740 &  1.40 &$-$2.32 &\nodata & 0.26                 &  0.14 &$-$0.02 &\nodata&$-$0.35  &high-res& GIL13\tablenotemark{b,c}\\
 Boo-34&  4840 &  1.60 &$-$3.10 &   0.55 &\nodata               &\nodata&\nodata&\nodata&  \nodata&med-res & NOR10\\ 

 Boo-41&  4750 &  1.60 &$-$1.93 &$-$0.65 &\nodata               &\nodata&\nodata&\nodata&  \nodata&med-res & NOR10\\
(=Boo24)& 4798 &  1.63 &$-$1.65 &$-$0.80 & 0.46\tablenotemark{a}&\nodata&\nodata&\nodata&  \nodata&med-res & LAI11\\
       &  4750 &  1.50 &$-$1.88 &\nodata & 0.50                 &  0.28 & 0.78  &\nodata& $-$0.39 &high-res& GIL13\tablenotemark{b,c}\\
 Boo-78&  4950 &  1.90 &$-$2.46 &$-$0.15 &\nodata               &\nodata&\nodata&\nodata&  \nodata&med-res & NOR10\\
 Boo-94&  4570 &  0.80 &$-$2.90 &$-$0.45 &\nodata               &\nodata&\nodata&\nodata&  \nodata&med-res & NOR10\\
       &  4600 &  0.50 &$-$2.95 & \nodata& 0.47                 &\nodata&  0.22 &\nodata&  \nodata&high-res& FEL09\\
       &  4560 &  0.80 &$-$2.94 & \nodata& 0.49                 &  0.30 &  0.26 &\nodata& $-$0.95 &high-res& GIL13\tablenotemark{b,c,d}\\ 
       &  4500 &  0.80 &$-$3.18 &$<$0.25 & 0.39                 &  0.46 &  0.55 &$-$2.16& $-$0.80 &high-res& ISH14\\
Boo-117&  4700 &  1.40 &$-$2.25 &$-$0.30 &\nodata               &\nodata&\nodata&\nodata&  \nodata&med-res & NOR10\\
(=Boo01)& 4716 &  1.65 &$-$2.34 &$-$0.50 & 0.12\tablenotemark{a}&\nodata&\nodata&\nodata&  \nodata&med-res & LAI11\\ 
       &  4600 &  1.00 &$-$2.29 & \nodata& 0.22                 &  0.29 &\nodata&\nodata& $-$0.46 &high-res& FEL09\\
       &  4725 &  1.40 &$-$2.18 & \nodata& 0.18                 &  0.20 &  0.14 &\nodata& $-$0.65 &high-res& GIL13\tablenotemark{b,c,d}\\
       &  4750 &  1.50 &$-$2.15 &$-$0.79 & 0.04                 &  0.01 &  0.44 &$-$1.12& $-$0.43 &high-res& ISH14\\
Boo-119&  4770 &  1.40 &$-$3.33 & \nodata& 1.04                 &  0.46 &  0.80 &\nodata& $-$1.00 &high-res& GIL13\tablenotemark{c,e}\\
(=Boo21)& 4775 &  1.48 &$-$3.79 &   2.20 & 0.27\tablenotemark{a}&\nodata&\nodata&\nodata&  \nodata&med-res & LAI11\\ 
Boo-121&  4630 &  1.10 &$-$2.37 &$-$0.25 &\nodata               &\nodata&\nodata&\nodata&  \nodata&med-res & NOR10\\
       &\nodata&\nodata&$-$2.44 &\nodata &  0.62                &  0.38 &\nodata&\nodata& $-$0.43 &high-res& FEL09\\
       &  4500 &  0.80 &$-$2.49 &$<-$0.24&  0.20                &  0.24 &  0.07 &\nodata& $-$0.63 &high-res& ISH14\tablenotemark{b,c}\\ 
Boo-127&  4670 &  1.40 &$-$2.08\tablenotemark{f} &$-$0.25 &\nodata               &\nodata&\nodata&\nodata&  \nodata&med-res & NOR10\\
       &  4600 &  1.00 &$-$2.03 &\nodata & 0.76                 &  0.02 &\nodata&\nodata&$-$ 0.64 &high-res& FEL09\\
       &  4685 &  1.40 &$-$2.01 &\nodata & 0.17                 &  0.16 &  0.20 &\nodata&$-$0.69  &high-res& GIL13 \\ 
       &  4750 &  1.60 &$-$1.92 &$-$0.77 & 0.11                 &  0.19 &  0.20 &$-$1.44& $-$0.87 &high-res& ISH14\\
       &  4765 &  1.35 &$-$1.99 &$-$0.15 &  0.41                &  0.24 &  0.33 &$-$1.20& $-$0.75 &high-res& this study\tablenotemark{b}\\
Boo-130&  4750 &  1.40 &$-$2.29 &$-$0.40 &\nodata               &\nodata&\nodata&\nodata&  \nodata&med-res & NOR10\\
       &  4775 &  1.40 &$-$2.32 & \nodata& 0.21                 &  0.19 &  0.17 &\nodata&$-$0.54  &high-res& GIL13\tablenotemark{b,c,d}\\
Boo-911&  4540 &  1.10 &$-$1.98 &$-$0.55 &\nodata               &\nodata&\nodata&\nodata&  \nodata&med-res & NOR10\\
       &\nodata&\nodata&$-$2.26 & \nodata& 0.11                 &  0.40 &\nodata&\nodata& $-$0.56 &high-res& FEL09\\
       &  4500 &  0.90 &$-$2.16 &$-$0.77 & 0.35                 &$-$0.01&  0.02 &\nodata& $-$0.64 &high-res& ISH14\tablenotemark{b}\\
Boo-980&  4890 &  1.70 &$-$3.09 &   0.00 &\nodata               &\nodata&\nodata&\nodata&  \nodata&med-res & NOR10\\
       &  4720 &  1.40 &$-$3.06 &$-$0.40 & 0.29                 &  0.30 &  0.43 &$-$0.95& $-$1.23 &high-res& this study\tablenotemark{b}\\
Boo-1069& 5050 &  2.20 &$-$2.51 &$-$0.35 &\nodata               &\nodata&\nodata&\nodata&  \nodata&med-res & NOR10\\
Boo-1137& 4710 &  1.20 &$-$3.66 &   0.20 &\nodata               &\nodata&\nodata&\nodata&  \nodata&med-res & NOR10\\
       &  4700 &  1.20 &$-$3.66 &   0.25 & 0.47                 &  0.52 &  0.52 &$-$1.42& $-$0.59 &high-res & NOR10b\tablenotemark{b}\\\hline\\
\multicolumn{12}{c}{Adopted abundances of high-resolution spectroscopic studies}\\\hline
 Boo-7 &\nodata&\nodata&$-$2.33 &$-$0.49(0.07\tablenotemark{g})   & 0.30         &  0.23 &\nodata&\nodata& $-$0.75 &high-res& FEL09,NOR10\\
 Boo-9 &  4750 &  1.40 &$-$2.64 &$-$0.58($-$0.02\tablenotemark{g})& 0.13         &  0.18 & 0.27  &\nodata& $-$0.89 &high-res& ISH14,NOR10\\
 Boo-33&  4740 &  1.40 &$-$2.32 &   0.29(0.74\tablenotemark{g})   & 0.26         &  0.14 &$-$0.02&\nodata& $-$0.35 &high-res& GIL13,NOR10\\
 Boo-41&  4750 &  1.50 &$-$1.88 &$-$0.78($-$0.24\tablenotemark{g})& 0.50         &  0.28 & 0.78  &\nodata& $-$0.39 &high-res& GIL13,NOR10\\
 Boo-94&  4560 &  0.80 &$-$2.94 &$-$0.44(0.31\tablenotemark{g})   & 0.49         &  0.30 &  0.26 &$-$2.43& $-$0.95 &high-res& GIL13,NOR10,ISH14\\
Boo-117&  4725 &  1.40 &$-$2.18 &$-$0.50(0.04\tablenotemark{g})   & 0.18         &  0.20 &  0.14 &$-$1.22& $-$0.65 &high-res& GIL13,NOR10,ISH14\\
Boo-119&  4770 &  1.40 &$-$3.33 &   1.85(2.16\tablenotemark{g})   & 1.04         &  0.46 &  0.80 &\nodata& $-$1.00 &high-res& GIL13,LAI11\\
Boo-121&  4500 &  0.80 &$-$2.49 &$-$0.13(0.62\tablenotemark{g})   & 0.20         &  0.24 &  0.07 &\nodata& $-$0.63 &high-res& ISH14,NOR10\\
Boo-127&  4765 &  1.35 &$-$1.99 &$-$0.15(0.37\tablenotemark{g})   & 0.41         &  0.24 &  0.33 &$-$1.20& $-$0.75 &high-res& FRE14\\
Boo-130&  4775 &  1.40 &$-$2.32 &$-$0.41(0.17\tablenotemark{g})   & 0.21         &  0.19 &  0.17 &\nodata& $-$0.54 &high-res& GIL13,NOR10\\
Boo-911&  4500 &  0.90 &$-$2.16 &$-$0.77($-$0.01\tablenotemark{g})& 0.35         &$-$0.01&  0.02 &\nodata& $-$0.64 &high-res& ISH14\\ 
Boo-980&  4720 &  1.40 &$-$3.06 &$-$0.40(0.13\tablenotemark{g})   & 0.29         &  0.30 &  0.43 &$-$0.95& $-$1.23 &high-res& FRE14\\
Boo-1137& 4700 &  1.20 &$-$3.66 &   0.25(0.90\tablenotemark{g})   & 0.47         &  0.52 &  0.52 &$-$1.42& $-$0.59 &high-res& NOR10b\\
\enddata 

\tablecomments{Note the different numbering schemes, e.g., Boo07 and
  Boo-7 do not refer to the same star. Different designations for the
  same star are indicated. References:
FEL09: \citet{feltzing09};  
NOR10: \citet{norris10_booseg};
NOR10b: \citet{norris10}; 
LAI11: \citet{lai11};
GIL13: \citet{gilmore13}; 
ISH14: \citet{ishigaki14}; 
FRE14: this study. 
``med-res'' refers to medium-resolution spectroscopy results,
``high-res'' refers to high-resolution spectroscopy results.}
\tablenotetext{a}{These values are [$\alpha$/Fe], not [Mg/Fe].}
\tablenotetext{b}{Abundances from this study have been adopted for this star.}
\tablenotetext{c}{[C/Fe] values of Norris et al. (2010), adjusted for the different [Fe/H] abundances, were used.}
\tablenotetext{d}{[Sr/Fe] values of Ishigaki et al. (2014), adjusted for the different [Fe/H] abundances, were used.}
\tablenotetext{e}{[C/Fe] values of Lai et al. (2011), adjusted for the different [Fe/H] abundances, was used.}
\tablenotetext{f}{[Fe/H] from Norris et al. (2010) is based on early analysis of a high-resolution UVES spectrum. The medium-resolution value is $\mbox{[Fe/H]} = -1.49$.}
\tablenotetext{g}{Final carbon abundances corrected for evolutionary status of the star (following Placco et al. 2014) is given in parenthesis.}
\end{deluxetable*}

\section{On the origin and evolution of Bo\"otes\,I}\label{signature}

 Using all available abundances of stars in Bo\"otes\,I, as
  described in the previous section, we now investigate the global
  chemical abundance signatures to characterize the origin and history
  of this ultra-faint dwarf galaxy.  In particular, we use four
  abundance criteria developed by \citet{frebel12} to assess to what
  extent chemical evolution has taken place in this system and how far
  removed Bo\"otes\,I might be from being a surviving first galaxy,
  like e.g., Segue\,1. The abundance criteria concern the a)
  metallicity distribution function (MDF), b) light and iron-peak
  element abundances, c) $\alpha$-elements, and d) neutron-capture
  elements. Under e) we also add a discussion on the role of carbon in
  these early systems. All of these are discussed in detail below,
  and applied to the body of abundance data presently available for
  Bo\"otes\,I .

\begin{figure}[!th]                                             
 \begin{center}                                                               
   \includegraphics[clip=true,width=8cm,bbllx=67, bblly=130,
     bburx=400, bbury=750]{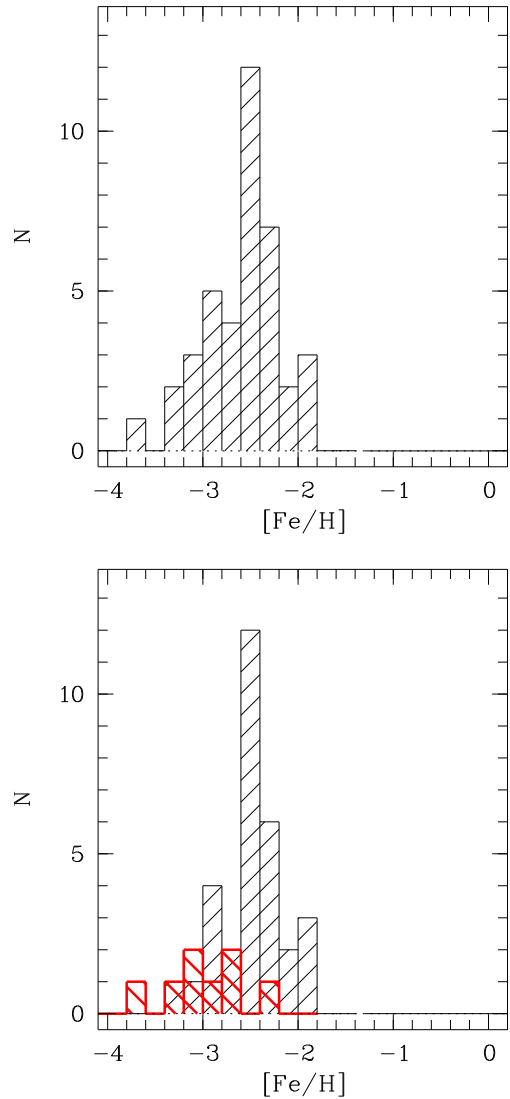} \figcaption{ \label{bootes_mdf}
     Metallicity distribution function for Bo\"otes\,I.  Note the lack
     of metal-rich stars above $\mbox{[Fe/H]} = -1.8$. Top panel:
     Complete sample of 39 stars. Bottom panel: red histogram depicts
     CEMP stars with $\mbox{[C/Fe]} >0.7$, black histogram the
     remaining non-CEMP stars. Additional discussions about the CEMP
     histogram in part e) of this section.}
 \end{center}                                                              
\end{figure}

\textbf{a) Iron abundance spread and metallicity distribution
  function.} This criterion stipulates that a large spread of 2-3\,dex
in [Fe/H] as well as a not too steep MDF shape (especially on the
low-metallicity side) would be signs of early inhomogeneous
mixing. The upper panel of Figure~\ref{bootes_mdf} shows the MDF of
Bo\"otes\,I as obtained from the sample of 39 stars that includes both
medium-resolution and high-resolution [Fe/H] abundances
(\citealt{norris10_booseg,lai11,feltzing09, norris10, gilmore13,
  ishigaki14}, this work). When available, a high-resolution value was
used for a given star.

A spread in [Fe/H] of 2\,dex is apparent, ranging from $\mbox{[Fe/H]}
\sim -3.8$ to $\mbox{[Fe/H]} \sim -1.8$. The lack of stars
$\mbox{[Fe/H]} > -1.8$ (when considering high-resolution abundances
where available) is typical for ultra-faint dwarf galaxies (e.g.,
\citealt{kirby08,ufs,frebel14}). The mean value for Bo\"otes\,I of
$\mbox{[Fe/H]} =-2.6$ based on this combined sample is in line with
previous results ($\mbox{[Fe/H]} =-2.3$, Belokurov et al. 2006;
$\mbox{[Fe/H]} =-2.5$, \citealt{munoz06}, \citealt{siegel06},
\citealt{martin07} and \citealt{norris_boo}; $\mbox{[Fe/H]} =-2.6$,
\citealt{lai11}).

From a theoretical viewpoint, chemical inhomogeneity likely resulted
in a large spread in [Fe/H] ranging from $\mbox{[Fe/H]} \sim-4.0$ up
to $\sim-1.0$ \citep{greif11}, without a very pronounced peak in the
[Fe/H] distribution. This behaviour is found in Segue\,1
\citep{frebel14} which now presumably contains only those first
Pop\,II stars.  In Bo\"otes\,I, a similar population of the earliest
Pop\,II stars could have formed, with their highest metallicity stars
having $\mbox{[Fe/H]} \sim-2.0$. Any additional Bo\"otes\,I stars
located at and near the peak region of the MDF might then be stars
from the second and/or later stellar generations formed from more
homogeneous gas enriched by any first/early Pop\,II supernovae. Still,
star formation must have been rapidly quenched to prevent stars with
$\mbox{[Fe/H]} >-1.8$ to form.

Regarding the MDF shape, there is a steady increase from
$\mbox{[Fe/H]} \sim -4.0$ to the distinct peak of the distribution at
$\mbox{[Fe/H]} \sim -2.5$. Then, there is a relatively sharp drop-off
towards higher metallicities suggesting that star formation was
extinguished rather abruptly, preventing the formation of stars with
$\mbox{[Fe/H]} >-1.8$. The same shape of the metallicity distribution
has also been found in other, more luminous dwarf galaxies (e.g.,
Draco and Carina, although they have higher peak metallicities; Fig~17
of \citealt{norris10}). Since this behavior is thus not unique to
Bo\"otes\,I we take it as a sign of chemical evolution (as opposed to
a limited chemical enrichment), although likely in its early phases given
the low peak metallicity.

\citet{lai11} explored the nature of the Bo\"otes\,I MDF with simple
chemical evolution models, following \citet{kirby11}. Taking into
account our slightly more populated MDF which also contains more stars
with high-resolution [Fe/H] measurements, it appears that their
leaky-box model with pre-enriched initial gas would best describe the
low metallicity tail of Bo\"otes\,I. That model begins with
pre-enriched  gas of low level of $\mbox{[Fe/H]}_{0}
\sim-4.0$ which is close to the assumption of pristine gas. However,
it would simultaneously underpredict the very tall peak region
of the MDF. We speculate that this could signify a subsequent
population of stars that formed after the initial generation emerged
from gas enriched by the first/early Pop\,II supernovae.

\textbf{b) Core-collapse supernova signatures} This criterion
stipulates that light elements (including iron-peak elements) observed
in early dwarf galaxy stars should match those of equivalent
metal-poor halo stars as core-collapse supernovae are believed to be
the progenitors of these elements. As can be seen in Figure~3, the
observed abundances of stars in Bo\"otes\,I are in good agreement with
metal-poor halo star abundances, in line with a fast enrichment in the
earliest dwarf galaxies.

\textbf{c) $\alpha$-element abundances: late-time star formation?}  This
criterion stipulates that the $\alpha$-elements (Mg, Si, Ca, Ti)
should show enhanced abundances of $\sim0.4$\,dex due to core-collapse
supernova as progenitors for all stars, even at high(er) metallicities
($\mbox{[Fe/H]}>-2.0$) since it implies that no stars formed after
(any) late-time enrichment of iron by supernovae\,Ia. This would
support that only one/few generations of early Pop\,II stars formed in
a system.

As noted by previous author (e.g., \citealt{gilmore13}), the
$\alpha$-element abundances of stars in Bo\"otes\,I do not provide a
clear picture. Consideration of the individual $\alpha$-elements,
e.g., in Figure~\ref{abund_lit}, shows that all stars have [Mg/Fe]
high-resolution abundance ratios in agreement with halo stars. For
[Ca/Fe] and [Ti/Fe], most Bo\"otes\,I stars have typical halo values
but there are also several outliers with abundance near the solar
ratio. However, in all cases, the 2-3 most metal-rich stars do not
have the lowest abundances but rather halo-like values. In fact, the
highest metallicity star (Boo-41) has one of the highest
$\alpha$-element abundances ($\mbox{[$\alpha$/Fe]} = 0.52$). This
supports Bo\"otes\,I containing those early Pop\,II stars, although
not exclusively.

\begin{figure}[!th]
 \begin{center}
   \includegraphics[clip=true,width=9.3cm,bbllx=150, bblly=115,
     bburx=450, bbury=495]{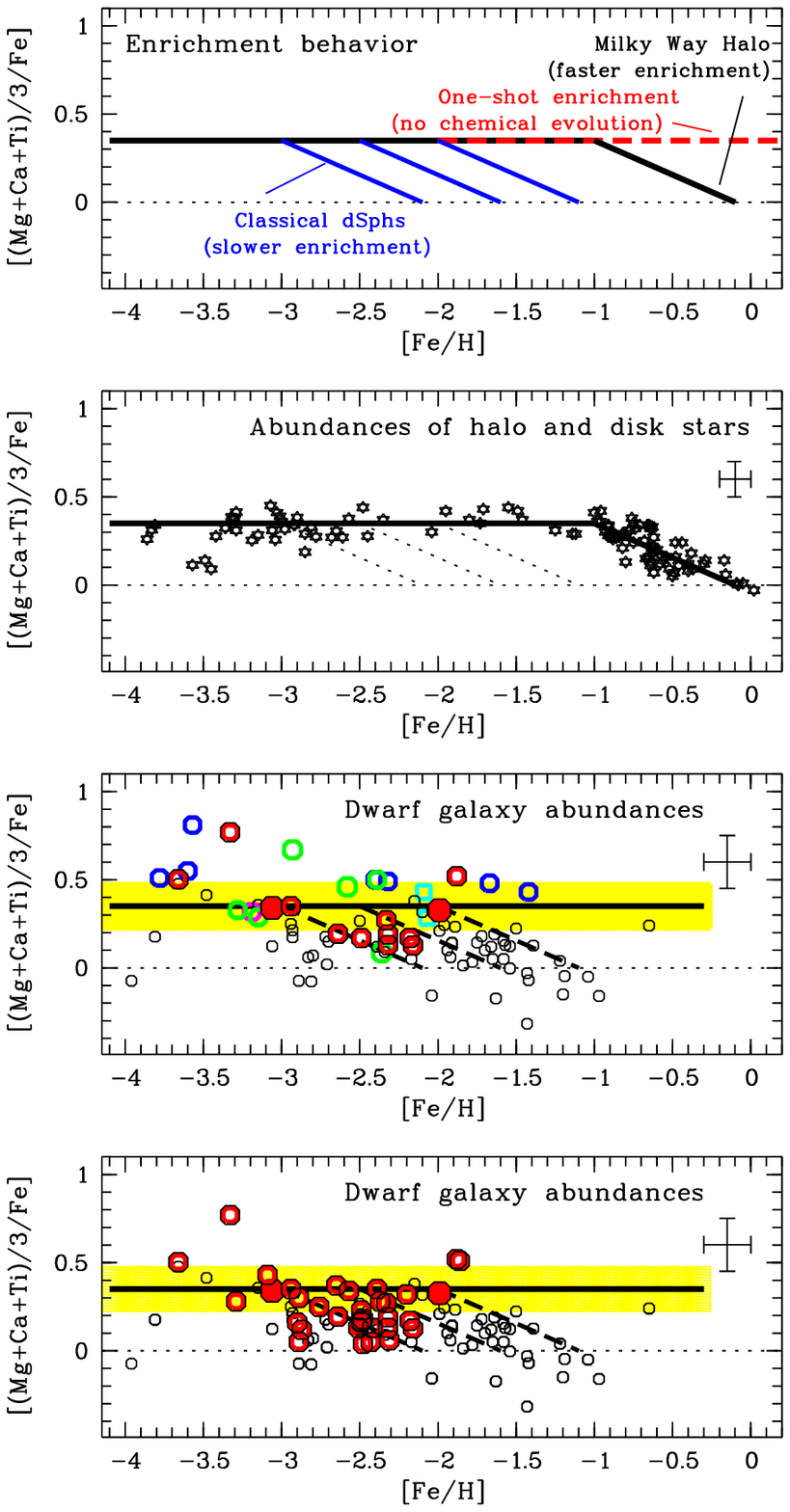}
   \figcaption{\scriptsize \label{oneshot} Combined Mg-Ca-Ti
     $\alpha$-element abundances as diagnostic of early star
     formation; adapted from \citet{frebel12}.  Top: Schematic
     representation of chemical enrichment in the [$\alpha$/Fe] vs
     [Fe/H] plane for different environments. The dotted line
     indicates the solar ratio. Second: High-resolution
     $\alpha$-abundances of metal-poor stars from Cayrel et al. (2004)
     (halo) and \citet{fulbright} (thin/thick disk). The diagonal
     dotted lines indicate the enrichment behavior of the dSph
     galaxies (see top panel). A representative uncertainty is shown.
     Third: The yellow shaded region around $\mbox{[$\alpha$/Fe]} =
     0.35$ depicts the predicted \citet{frebel12} one-shot enrichment
     behavior (with a 0.15\,dex observational uncertainty) reflecting
     massive core-collapse supernova enrichment. Several evolutionary
     paths are indicated with dashed lines.  High-resolution
     $\alpha$-abundances of metal-poor stars are shown; For
     ultra-faint dwarf galaxies: Bo\"otes I stars, this study
     (\textit{filled red circles}) Bo\"otes I, literature
     (\textit{open red circles}; see text for references), Segue\,1
     (\textit{blue circles}; \citealt{norris10_seg,frebel14}); Ursa
     Major II and Coma Berenices (\textit{green circles};
     \citealt{frebel10}); Leo\,IV (\textit{pink circle};
     \citealt{simon11}); Hercules (\textit{cyan squares};
     \citealt{koch_her}). For classical dSphs (\textit{small open
       black circles}; see \citealt{frebel12} and references therein).
Bottom: Same as above but with only the Bo\"otes\,I stars that include
the [$\alpha$/Fe] measurements of \citet{lai11}.
}
 \end{center}
\end{figure}

The combined $\alpha$-element abundances are then helpful to assess
whether Bo\"otes\,I experienced late time, extended star formation as
evidenced by low(er) $\alpha$-abundances. This is illustrated in
Figure~\ref{oneshot}. In the bottom two panels, we show the
Bo\"otes\,I abundances together with abundances of other ultra-faint
dwarf galaxies as well as those of stars in the classical dwarf
spheroidal galaxies. Several Bo\"otes\,I stars have abundances
resembling those of the halo, Segue\,1 and other ultra-faint dwarf
galaxy stars while many others have values very similar to those of
the classical dwarf galaxy stars (21 have $\mbox{[$\alpha$/Fe]} <
0.3$ and 16 $\mbox{[$\alpha$/Fe]} < 0.2$).

It thus appears like Bo\"otes\,I is a system that was assembled from
or absorbed one (or a few) smaller building block-type objects (e.g.,
minihalos) like Segue\,1 before forming more stars and experiencing a
chemical evolution (as opposed to just chemical enrichment as, e.g.,
in Segue\,1) that would eventually lead to stars with low
$\alpha$-abundances. Such a two-population scenario is qualitatively
in line what was found based on the CEMP and non-CEMP star MDF
components of Bo\"otes\,I. More precise $\alpha$-element abundances
will be needed for all stars (e.g., recall that Lai et al. 2011 only
provided a combined $\alpha$-measurement) to conclusively determine
whether any late time enrichment by supernova type\.Ia did indeed
occur in Bo\"otes\,I. This would indicate extended star formation
possibly due to an gas accretion from other clouds or cloud fragments
in the system or an early merger with a different system that would
have brought in extra gas \citep{smith15}.

\textbf{d) Low and unusual neutron-capture element abundances: evidence
  for one progenitor generation?}  This criterion stipulates that 
abundances of heavy neutron-capture elements (e.g., Sr and Ba) should
be very low since it implies that no stars formed after late(r) time
enrichment associated with the s-process operating in AGB stars. Small
amounts of neutron-capture elements, primarily with [Sr/Ba] ratios
characteristic of the r-process (occurring in supernovae) could, however,
be present in the system without implying late time star formation.

We have already discussed that the neutron-capture elements Sr and Ba are
of low abundance in Bo\"otes\,I (Figure~5). Specifically, the Sr
abundances are not like those of halo stars but up to $\sim1$\,dex
lower. Ba does roughly follow the halo star abundance trend but all
Bo\"otes\,I stars still have abundances that place them at the lower
edge of the region covered by halo data. In addition, the spread in
neutron-capture element abundances in Bo\"otes\,I, as evidenced by
only four stars, is about 1\,dex. This precludes overarching
statements regarding the exact neutron-capture process responsible for
this material.  The stars with $\mbox{[Sr/Ba]}\sim-0.5$ ratio
technically fall in the regime of the r-process (we note that most
Segue\,1 stars are also near this value). The s-process is
characterized by $\mbox{[Sr/Ba]}<-1.0$ and indeed one Bo\"otes\,I star
has such low [Sr/Ba] value. Overall, this is consistent with
Bo\"otes\,I being an old system that formed its stars early and
relatively quickly, before the onset of any regular AGB star-based
s-process element enrichment. Consequently, fast neutron-capture
enrichment might have occurred by just one of the earliest generations
of massive stars in the system.

\begin{figure}[!tb]
  \begin{center}
   \includegraphics[clip=true,width=9cm,bbllx=50, bblly=100,
     bburx=405, bbury=768]{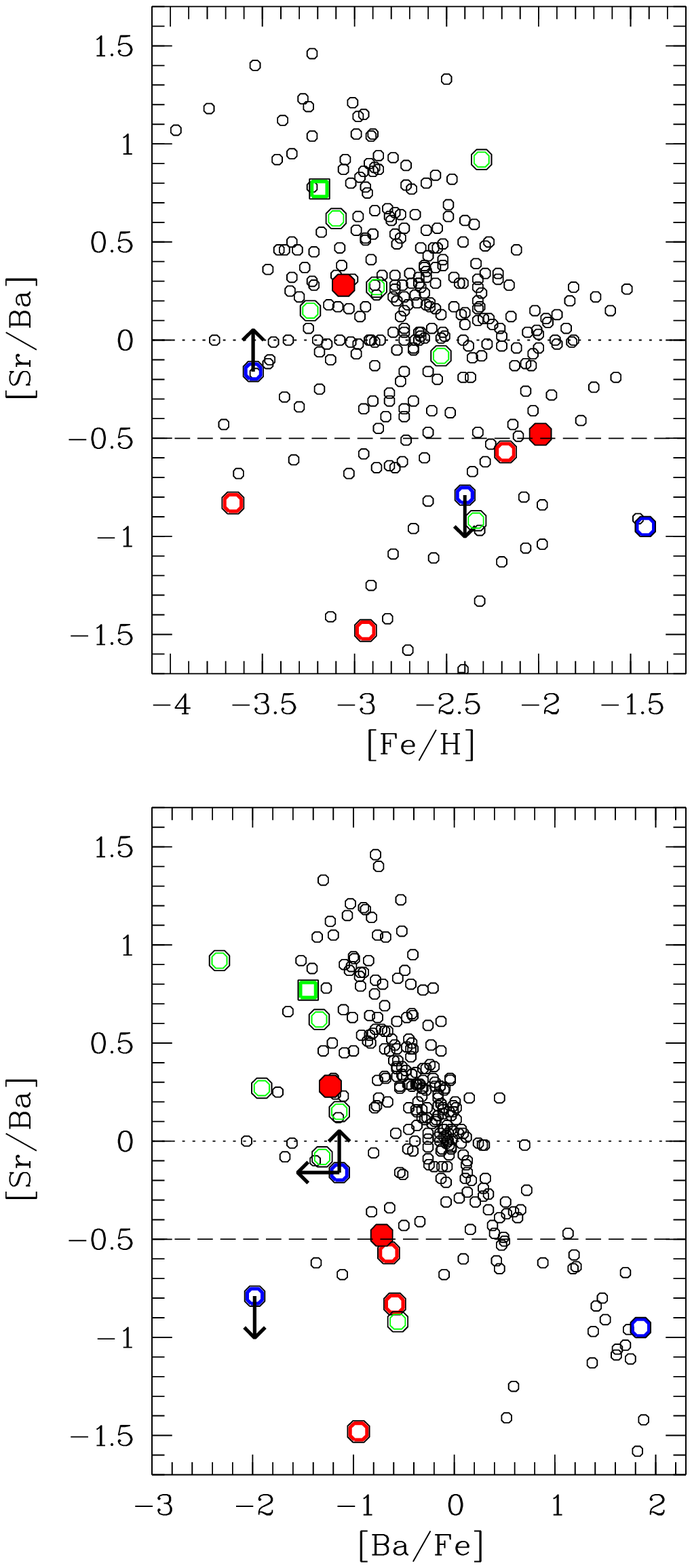}
   \figcaption{\scriptsize \label{srba_plot} [Sr/Ba] abundance ratios
     as a function of [Fe/H] (top panel) and [Ba/Fe] (bottom panel) of
     our Bo\"otes\,I stars (\textit{filled red circles}) and
     literature Bo\"otes\,I stars (\textit{open red circles}; here the
     \citealt{ishigaki14} values were used for internal consistency) in
     comparison with those of other ultra faint dwarf galaxy stars in
     Segue\,1 (\textit{blue circles};
     \citealt{frebel14,norris10_seg}), UMa\,II, ComBer, Leo\,IV
     (\textit{green circles}; \citealt{ufs, leo4}), and the galactic
     halo (\textit{black open circles}; \citealt{heresII} and
     \citealt{yong13_II}). }
  \end{center}
 \end{figure}

In Figure~\ref{srba_plot}, there exists no obvious correlation between
[Sr/Ba] and [Fe/H]. When plotting [Sr/Ba] vs. [Ba/Fe] (see bottom
panel). However, a clear trend emerges. Interestingly, the halo stars
form a distinct branch while the various dwarf galaxy stars form
something that resembles a second sequence below the halo branch. The
few Bo\"otes\,I stars with measured Sr abundances follow this trend
set by the UMa\,II, ComBer, Leo\,IV, and three Segue\,1 stars. Some
halo stars do exhibit similar neutron-capture element characteristics
to those of the dwarf galaxy stars. This might indicate that these
halo stars are stars that originated in small dwarf galaxies similar
to those in the surviving ultra-faint dwarfs. While more data for
ultra-faint dwarf galaxy stars is clearly needed to more firmly
establish this second sequence, perhaps also more halo data could help
shape up any of the differences between the halo star branch and the
dwarf galaxy sequence even more clearly.

Using neutron-capture element abundances is clearly a promising way
path to learn not only about the nature of the host dwarf galaxy but
also about the origin of halo stars. Lighter element abundances have
been repeatedly shown to all agree with each other for halo and dwarf
galaxy stars making it more challenging to use them as a
discriminator.

Alternatively, the distinct patterns of halo stars and dwarf galaxy
stars could originate from highly inhomogeneous enrichment of their
respective birth gas clouds. However, while this would explain the
large spreads in [element/H] abundance ratios such as [Fe/H] or
[Sr/H], [element/element] ratios should in principle be hardly
affected \citep{frebel12}. Indeed, the lighter elements show tight
[element/Fe] ratios and support this possibility. In turn, the large
spread in [Sr/Ba] might indicate that the two elements are not
produced with a strong correlation between them as the
fusion-produced elements are, or that metal mixing did not play a
primary role. 

\textbf{e) Carbon enrichment: prevalent at early
  times?}\label{sec:carbon} After applying carbon abundance
corrections according to stellar evolutionary status (see
Table~\ref{lit_abund}), there are eight CEMP stars in Bo\"otes\,I with
metallicities below $\mbox{[Fe/H]}\lesssim-2.6$. Incidentally, the
CEMP stars are all CEMP-no stars and rather equally distributed among
the lower metallicity half of all Bo\"otes\,I stars. All stars have
[C/H] values higher than $\mbox{[C/H]}=-3.4$ placing them well above
the critical carbon and oxygen abundance criterion of $D_{\rm{trans}}
= \log(10^{\rm{[C/H]}} +0.3\times10^{\rm{[O/H]}})=-3.5$.  that may
have facilitated their formation from carbon and oxygen enriched gas
in the early universe \citep{dtrans}. Carbon must thus have played an
important role in the early evolution of Bo\"otes\,I.

[C/H] ratios of the Bo\"otes\,I sample do not show a clear trend with
either [Fe/H] or $\log g$, although the sample does follow the overall
carbon abundance range given by halo stars (see Figure~\ref{cfe}, also
for references). The CEMP stars in Bo\"otes\,I increasingly appear all
at lower metallicities, at $\mbox{[Fe/H]}\le-2.6$. Generally, the
high-carbon outliers may indicate that the production of carbon and
iron and their subsequent mixing into the gas has been decoupled or
that CEMP and non-CEMP signatures have been produced at different
sites. For a more extended discussion on the origin of CEMP stars, we
refer the reader to \citet{norris13_IV} or \citet{fn15}.

\citet{cooke14} suggested a minihalo environment as a site for CEMP
star formation with faint fall-back first supernovae (e.g.,
\citealt{UmedaNomotoNature}) responsible for large quantities of
lighter elements such as carbon but for less or none of the heavier
ones like iron. The bottom panel of Figure~\ref{bootes_mdf}  shows the separate MDFs
for CEMP stars (red) and non-CEMP stars (black). The CEMP stars
cluster at low metallicities albeit with some spread in [Fe/H]. But
chemical inhomogeneity could possibly account for this (e.g.,
\citealt{frebel12}). These CEMP stars thus might have formed in a
minihalo environment that would later become Bo\"otes\,I or part of
Bo\"otes\,I.

The non-CEMP star MDF lacks the pronounced low-metallicity tail and an
average metallicity of $\mbox{[Fe/H]}=-2.5$. This leaves the MDF to be
more like the one predicted by the leaky box chemical evolution model
with extra gas \citep{lai11,kirby11}. This incorporates an additional
reservoir of gas available to the galaxy from which to form stars. A
merger of said minihalo with another early halo (presumably prior to
it forming any stars) could have provided this gas. These findings
build in some way on the conclusions of \citet{gilmore13} who suggest
that the ``CEMP-no'' stars (our ``CEMP stars'') formed rapidly and
prior to the ``normal branch'' (our ``non-CEMP'' stars).

More generally, massive rotating Pop\,III ``spinstars''
\citep{meynet06,chiappini11} could have dominated the enrichment of
Bo\"otes\,I at the earliest times by providing large amounts on carbon
to the star forming gas prior to their explosions. Fall-back
supernovae \citep{UmedaNomotoNature} could have been equally
responsible for high [C/Fe] ratios in environments more massive than
minihalos. Regardless, the earliest low-mass stars would likely show a
variety of [C/Fe] ratios, and a clear trend would only emerge with the
onset of more regular Pop\,II supernovae and extended star formation.

\section{Conclusion}\label{sec:conc}

We have studied the stellar chemical abundances of Bo\"otes\,I with
high-resolution spectroscopy, including a previously unobserved member
star. Considering all available abundance data, we examined
Bo\"otes\,I in light of several criteria for identifying the earliest galaxies. The
$\alpha$- and carbon abundances suggest two potential populations,
although more precise abundances are required to draw firm
conclusions. Considerations of these chemical signatures alone suggests that
Bo\"otes\,I is not a surviving first galaxy but an already somewhat
assembled system that could have been built up from a minihalo (a
first galaxy) in which the CEMP stars formed \citep{cooke14}, and then
merged with another metal-free or metal-poor halo (or large gas clouds
or filaments) that resulted in an injection of fresh gas
triggering star formation and leading to a second population of
stars. Both populations show [Fe/H] spreads of $>1$\,dex which is in
line with predictions of chemical inhomogeneity (e.g.,
\citealt{frebel12}) and expected for such early star formation.

Similar results have also been derived by \citet{romano15} who modeled
the formation and chemical evolution of Bo\"otes\,I. They conclude
that Bo\"otes\,I formed from accretion of a baryonic mass M$_b \sim
10^{7}$\,M$_{\odot}$ of gas over a very short time scale of order
50\,Myr. This is also in agreement with what was already qualitatively
predicted by \citet{gilmore13} who found that the CEMP stars formed
rapidly and prior to the non-CEMP stars.

All these findings are highly consistent with results based on deep
HST photometry. \citet{brown14b} found Bo\"otes\,I to currently
consist of two ancient stellar populations of 13.4\,Gyr (containing
$3\%$ of stars) and 13.3\,Gyr (97$\%$ of stars). We note,
  though, that these ages are relative to the globular cluster M92,
  and assuming its age to be 13.7\,Gyr. Nevertheless, it is apparent
  that all the dwarf galaxies contain one ancient dominant stellar
  population plus a small age spread.  The two-population scenario
follows \citet{koposov11} who also found that Bo\"otes\,I is best
described with two components, based on a kinematic analysis. They
identified a ``colder'' ($\sim3$\,km\,s$^{-1}$) component (containing 70\% of stars) and a
``hotter'' $\sim9$\,km\,s$^{-1}$ component (30\% of stars).  In addition, they suggested
that the colder component have ``an extremely extended, very
metal-poor, low-velocity dispersion component''. The chemical
abundances and the MDF available for Bo\"otes\,I stars clearly show
the existence of such a low-metallicity tail (see
Figure~\ref{bootes_mdf}) that is mainly composed of CEMP stars.

In summary, evidence based on multiple independent approaches suggest
Bo\"otes\,I to be an ancient system with two stellar populations of
which at least the more metal-poor one likely consisted of the first
low-mass stars that formed in its progenitor halo(s). This makes
Bo\"otes\,I an assembled system with a more complex evolution than
what would be expected of a surviving galactic building block, such as
Segue\,1. Nevertheless, Bo\"otes\,I is likely one of the earliest
galaxies to have formed. Not just for its age of $\sim13.4$\,Gyr
\citep{brown14b} but also because we seem to be witnessing what is the
product of two systems, one minihalo and one larger halo which merged
to form Bo\"otes\,I. To some extent this outcome is already indicated
by its larger luminosity of nearly $\sim10^5$\,L$_{\odot}$, compared
to e.g., Segue 1 with only about $\sim10^3$\,L$_{\odot}$. This leaves
the question as to whether an average metallicity of an early galaxy is a
useful quantify if more than one population contributes stars in
different metallicity ranges. Ideally, one could cleanly disentangle
the population(s) to report separate results. This might work for
Bo\"otes\,I but will surely become impossible for more complex, more
luminous systems. 

More ultra-faint galaxies need to be extensively studied with
photometry and spectroscopy from a chemical and kinematic point of
view to assess the nature of as many as possible of the faint dwarfs currently
known. Only then can be better understand the early evolution of
galaxies and the details of the associated star formation, chemical
enrichment and overall growth mechanisms and timescales of these
ancient systems.

\acknowledgements{A.F. is supported by NSF CAREER grant
  AST-1255160. She also acknowledges support from the Silverman (1968)
  Family Career Development Professorship. Studies at RSAA, ANU, of
  the Galaxy's most metal-poor stars and ultra-faint satellite systems
  are supported by Australian Research Council grants DP0663562,
  DP0984924, DP120100475, and DP150100862 which J.E.N. gratefully
  acknowledges.  This work made use of the NASA's Astrophysics Data
  System Bibliographic Services.}

\textit{Facilities:} \facility{Magellan-Clay (MIKE)}


\end{document}